\documentclass[aps,pra,twocolumn,reprint,floatfix, superscriptaddress]{revtex4-2}

\usepackage[T1]{fontenc}
\usepackage[utf8]{inputenc}
\usepackage{amssymb}

\usepackage{subfigure}
\usepackage{psfrag,graphicx}
\usepackage{pict2e}
\usepackage{dcolumn}
\usepackage{amsmath,amssymb,braket}
\usepackage{bm}
\usepackage{color}
\usepackage{latexsym}
\usepackage{epstopdf}
\usepackage{color}
\usepackage[english]{babel}
\UseRawInputEncoding
\usepackage{amsfonts}
\usepackage{amsmath}
\usepackage{bbm}
\usepackage{bm}%
\usepackage{natbib}
\usepackage{bbold}
\usepackage{float}
\usepackage{enumerate}

\usepackage{xr}

\definecolor{myblue}{rgb}{0.2,0.2,0.8}
\definecolor{myzard}{cmyk}{0,0,0.05,0}
\definecolor{mywhite}{rgb}{1,1,1}
\definecolor{mywhite}{rgb}{1,1,1}
\definecolor{myred}{rgb}{1,0.,0.3}
\usepackage[colorlinks=true,citecolor=myblue,linkcolor=myred]{hyperref}

\definecolor{mygrey}{gray}{0.35}
\definecolor{myblue}{rgb}{0.2,0.2,0.8}
\definecolor{myzard}{cmyk}{0,0,0.05,0}
\definecolor{mywhite}{rgb}{1,1,1}
\definecolor{mywhite}{rgb}{1,1,1}
\definecolor{myred}{rgb}{1,0.,0.3}

\def\be{\begin{equation}}
\def\ee{\end{equation}}
\def\ba{\begin{align}}
\def\enda{\end{align}}
\def\bi{\begin{itemize}}
\def\ei{\end{itemize}}

\def\beq{\begin{equation}}
\def\beq{\begin{equation}}
\def\eeq{\end{equation}}

\def\kk{{\textbf k}}

\def\rr{{\textbf r}}

\begin{document}

\title{Bound states in the continuum in subwavelength emitter arrays}

\author{Mar\'ia Blanco de Paz}
\affiliation{Donostia International Physics Center, 20018 Donostia-San Sebasti\'an, Spain}
\affiliation{Instituto de Telecomunica\c c\~oes, Instituto Superior Tecnico-University of Lisbon, Avenida Rovisco Pais 1, Lisboa, 1049-001 Portugal}

\author{Paloma A. Huidobro}
\email{p.arroyo-huidobro@uam.es}
\affiliation{Instituto de Telecomunica\c c\~oes, Instituto Superior Tecnico-University of Lisbon, Avenida Rovisco Pais 1, Lisboa, 1049-001 Portugal}
\affiliation{Departamento de F\'{i}sica Te\'orica de la Materia Condensada
, Universidad Aut\'onoma de Madrid, E-28049 Madrid, Spain}

\begin{abstract}
Ordered lattices of emitters with subwavelength periodicities support unconventional  forms of light-matter interactions arising from collective effects.  
Here, we propose the realization and control of subradiant optical states within the radiation continuum in two-dimensional lattices. We show how bound states in the continuum (BICs) which are completely decoupled from radiative states emerge in non-Bravais lattices of emitters. Symmetry breaking results in quasi-BICs with greatly extended lifetimes, which can be exploited for quantum information storage. The analytical derivation of a generalized effective polarizability tensor allows us to study the optical response of these arrays. We discuss how thanks to the quasi-BICs, a rich phenomenology takes place in the reflectivity spectrum, with asymmetric Fano resonances and an electromagnetically induced transparency window. Finally, we exploit these lattices as quantum metasurfaces acting as efficient light polarizers. 

\end{abstract}

\maketitle


The spontaneous decay of quantum emitters through coupling to radiative modes is an intrinsic quantum property of the emitters and represents a major source of decoherence in quantum photonic platforms. The suppression of this loss process is a key building block of quantum photonic technologies~\cite{bekensteinQuantum2020,NatPhot2021,ExperimentalNat2001} and can be achieved through preparation of photons in dark optical states~\cite{PhysRevA.78.053816,PhysRevLett.122.093601}. Dark, or subradiant, states emerge due to destructive interference arising from collective interactions between several quantum emitters \cite{PhysRev.93.99}, and provide a reduced coupling to radiative modes. This results in extended lifetimes that can improve the storage and processing of quantum information. 
Subradiant optical states have been observed in atom clouds in vacuum~\cite{PhysRevX.11.021031,PhysRevLett.125.213602,PhysRevLett.116.083601,PhysRevA.76.053814,science.1217901} or near an optical fibre~\cite{NatCommnanofiber17}, as well as with artificial atoms such as superconducting qubits~\cite{NatQED19,NatPhysQED22}. 

Recently, structured arrays of quantum emitters have also been proposed as a platform to realize subradiant states. These consist of periodic arrays of emitters with subwavelength periodicity, where coherent dipole-dipole interactions between all the emitters in the lattice give rise to cooperative effects that strongly modify the optical properties of the quantum emitters \cite{jenkins2012controlled,olmos2013longrange,bettles2016enhanced,schilder2016polaritonic,ruostekoski2016emergence,shahmoon2017cooperative,zhou2017optical}. These arrays support subradiant guided modes that only radiate due to the finite size of the system  and can be harnessed for selectively improving radiation in a given desired channel~ \cite{asenjo-garciaExponential2017}, for mediating non-trivial emitter-emitter interactions when additional atoms are placed nearby~\cite{Masson2020,Patti2021,Brechtelsbauer2020,Castells-Graells2021AtomicDimers,Fernandez-Fernandez2021TunableMetasurfaces,de2022manipulating} or for generating topological edge modes~\cite{perczel17a,bettles17a}. Periodic atomic arrays with hundreds of atoms have been experimentally realised  by means of optical tweezers~\cite{synthetic_nature,PhysRevLett.128.223202, PhysRevX.9.011057, natphys2021} or optical lattices~\cite{science.aaz6801,bloch08}, and an atomic optical mirror based on collective interactions has already been demonstrated \cite{ruiSubradiant2020,srakaew2022subwavelength}. 

On the other hand, subradiant modes can also arise within the radiation continuum. For instance, cooperative resonances can result in a reduced collective decay rate compared with that of an isolated quantum emitter \cite{PhysRevLett.117.243601} and magnetic responses can be created at optical frequencies ~\cite{alaeeQuantum2020,ballantineOptical2020}.
Here, we propose the realisation of completely dark states within the radiation continuum in subwavelength emitter arrays. To do so we exploit the concept of bound states in the continuum (BICs). While optical BICs have energies embedded in the radiation continuum, that is, are above the light-line, they cannot couple to radiative modes and  thus are completely dark. 
Instead of propagating, photons in a BIC remain bound to the structure and can serve as storage of quantum information.
BICs were first proposed in quantum mechanics as states that localize in space while having an energy higher than the potential well that confines them~\cite{bic29}, and then experimentally verified and analyzed in an 
acoustic system~\cite{PARKER196662,PARKER1967330}. Over the past years, BICs have been generalised to different wave physics scenarios~\cite{bicreview}, prominently in optics due to their potential for nanophotonic applications such as optical sensing~\cite{sensing1,sensing2} or lasing~\cite{APLlasing,APLlasing2,science.lasing1,science.lasing2,natlasing,Hirose2014WattclassHH,PhysRevLett.129.173901}.

\begin{figure}[ht]
  \centering
  \includegraphics[width=\columnwidth]{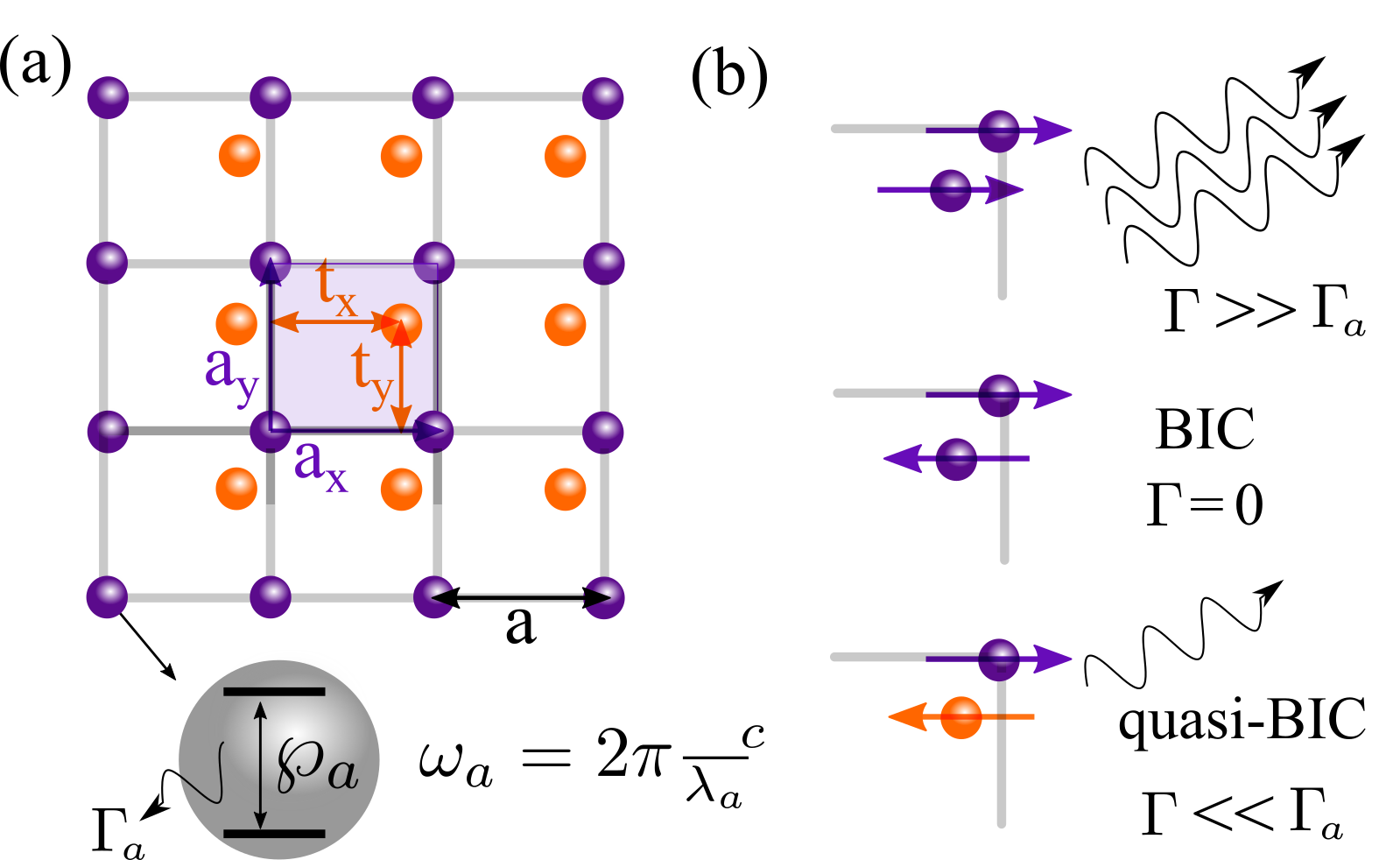}
  \caption{Subwavelength emitter arrays arranged in a square lattice. (a) Sketch of a lattice with a basis of 2. Two sets of emitters with resonance frequencies $\omega_{a,i}$ ($i=1,2$) are arranged in a square lattice of periodicity $a$ and lattice vectors $\mathbf{R}_1=\mathbf{a}_x=a\mathbf{x}$ and $\mathbf{R}_2=\mathbf{a}_y=a\mathbf{y}$. The basis vector is $\mathbf{T} = (t_x,t_y)$. (b) Optical response of arrays with a single emitter per unit cell. The contour plot shows reflectivity as function of frequency detuning $\delta=\omega-\omega_a$ normalised to the individual decay rate $\Gamma_a$ and period array normalised to the emitter's resonance wavelength $\lambda_a$. The cooperative shift $\Delta$ is plotted as green dashed line, highlighting how reflectivity peaks when $\delta=\Delta$. The lower panel depicts reflection and transmission as a function of period array for $\delta=0$ (marked by a gray line in the contour plots).  }
  \label{fig1Sketch}
\end{figure}

In order to generate BICs in subwavelength emitter arrays, we consider a two-dimensional non-Bravais lattice of emitters (see Fig.\ref{fig1Sketch}). Each lattice site contains one emitter, which in the single excitation approximation, can be modelled as two level systems with resonance wavelengths $\lambda_{a,i}$ and frequencies $\omega_{a,i}=2\pi c/\lambda_{a,i}$. The emitters' positions are given by $\mathbf{R}_{n}^{\mu} = \mathbf{R}_\mu + \mathbf{T}_n $, with $\{\mathbf{R}_n\}$ giving the positions of all unit cells ($n=1,\cdots, N$) and $\{\mathbf{T}_\mu\}$ being the basis vectors of the sublattices ($\mu=1,\cdots, M$), which join the origin of the unit cell with each of the emitters in the basis. The dynamics of the emitters can be described through an effective non-Hermitian Hamiltonian (see S.M.)~\cite{asenjo-garciaExponential2017,perczelPhotonic2017}:
\begin{align}
    \frac{H}{\hbar} &= \sum_{j=1}^{N_A}\left(\omega_j-i\frac{\Gamma_j}{2}\right) \sigma_{ee}^j  \nonumber \\ 
    &  -  \sum_{\substack{j=1 \\ i\neq j}}^{N_A} \frac{\omega_i^2}{\epsilon_0 c} \left[ \boldsymbol{\wp}^*_{i} \cdot \mathbf{G}_{ij}\cdot \boldsymbol{\wp}_j \right]\sigma_{eg}^i\sigma_{ge}^j\,.
\label{eq:H_eff}
\end{align}
Here, $N_A$ is the total number of emitters in the array, $\sigma^j_{ge,eg}$ are the atomic transition operators between the ground and excited states, $\Gamma_i=|\boldsymbol{\wp_i}|^2\omega_i^3/(3\pi\hbar c^3)$ are the spontaneous decay rate of the emitters and $\boldsymbol{\wp_i} $ the dipole moments. The second term in the Hamiltonian represents the photon-mediated interactions among all the emitters in the array. This is dictated by the Green's dyadic, $\mathbf{G}_{ij}=\mathbf{G}(|\rr_i-\rr_j|)$, with 
$\mathbf{G}(\rr)= \frac{1}{4\pi}\left[\mathbb{1}+\nabla\otimes\nabla/k^2\right]e^{ik |\rr|}/|\rr|$ and $k=\omega/c$, which includes short, medium and long range interactions. 
For infinite periodic lattices, the eigenstates of the Hamiltonian (Eq.~\ref{eq:H_eff}) are Bloch modes, $S^\dagger_\kk=\frac{1}{\sqrt{N}}\sum_{n=1}^N\sum_{\mu=1}^M \sigma^{n,\mu}_{eg} e^{i\kk\cdot\mathbf{r}_n}$, with $\kk$ the Bloch wavevector in the plane of the array, and where we label the atomic operators with lattice and sublattice indeces. The eigen-states and eigen-energies satisfy $H |\Psi_\mathbf{k}\rangle = \hbar \left(\omega_k - i \frac{\gamma_k}{2}\right)  |\Psi_\mathbf{k}\rangle$, and are thus obtained from the diagonalisation of a $3M\times3M$ matrix with elements,
\begin{align}
    \hat{\mathbf{N}}^{\mu\nu} = \left(\omega^\mu - i \frac{\Gamma_a}{2}\right) \mathbbm{1}\delta_{\mu\nu}  + \left(\hat{\Delta}^{\mu\nu} (\mathbf{k}) - i \frac{\hat{\Gamma}^{\mu\nu} (\mathbf{k})}{2}\right), 
\end{align}
Here, $\mathbbm{1} $ is the $3\times 3$ identity matrix and we have introduced the tensors 
\begin{align}
    \hat{\Delta}^{\mu\nu} (\mathbf{k}) &=  -\frac{3\pi\Gamma_a c  }{\omega_a} \textrm{Re}\left[\hat{\boldsymbol{\wp}}^{*} \cdot \hat{\mathbf{S}}^{\mu\nu}(\mathbf{k}_{||},\omega_a) \cdot \hat{\boldsymbol{\wp}}  \right] \\
    \hat{\Gamma}^{\mu\nu} (\mathbf{k}) &=  \frac{6\pi\Gamma_a c  }{\omega_a} \textrm{Im}\left[\hat{\boldsymbol{\wp}}^{*} \cdot \hat{\mathbf{S}}^{\mu\nu}(\mathbf{k}_{||},\omega_a) \cdot \hat{\boldsymbol{\wp}}  \right]. 
\end{align}
with $\hat{\boldsymbol{\wp}}=\boldsymbol{\wp}/|\boldsymbol{\wp}|$ and  $\hat{\mathbf{S}}^{\mu\nu}(\omega,\mathbf{k}_{||})$ standing for the Fourier transform of the dipole-dipole interaction. This is the lattice sum, 
\begin{equation} \label{eq:latticesums}
    \hat{\mathbf{S}}^{\mu\nu}(\mathbf{k}_{||}) =  \sideset{}{'}\sum_{m=1}^{N}  \hat{\mathbf{G}}(\mathbf{R}_{m} + \mathbf{T}^\nu -\mathbf{T}^\mu) e^{-i\mathbf{k}_{||} \mathbf{R}_{m}}\,
\end{equation}
where the primed symbol indicates that the sum excludes the term $m=1$ when $\mu=\nu$ \footnote{The lattice sums are slowly convergent due to the long range interactions in the Green's dyadic and we employ the Ewald method to perform them efficiently~\cite{ewald1921,Linton2010}.}.

\begin{figure}[ht]
  \centering
  \includegraphics[width=0.99\columnwidth]{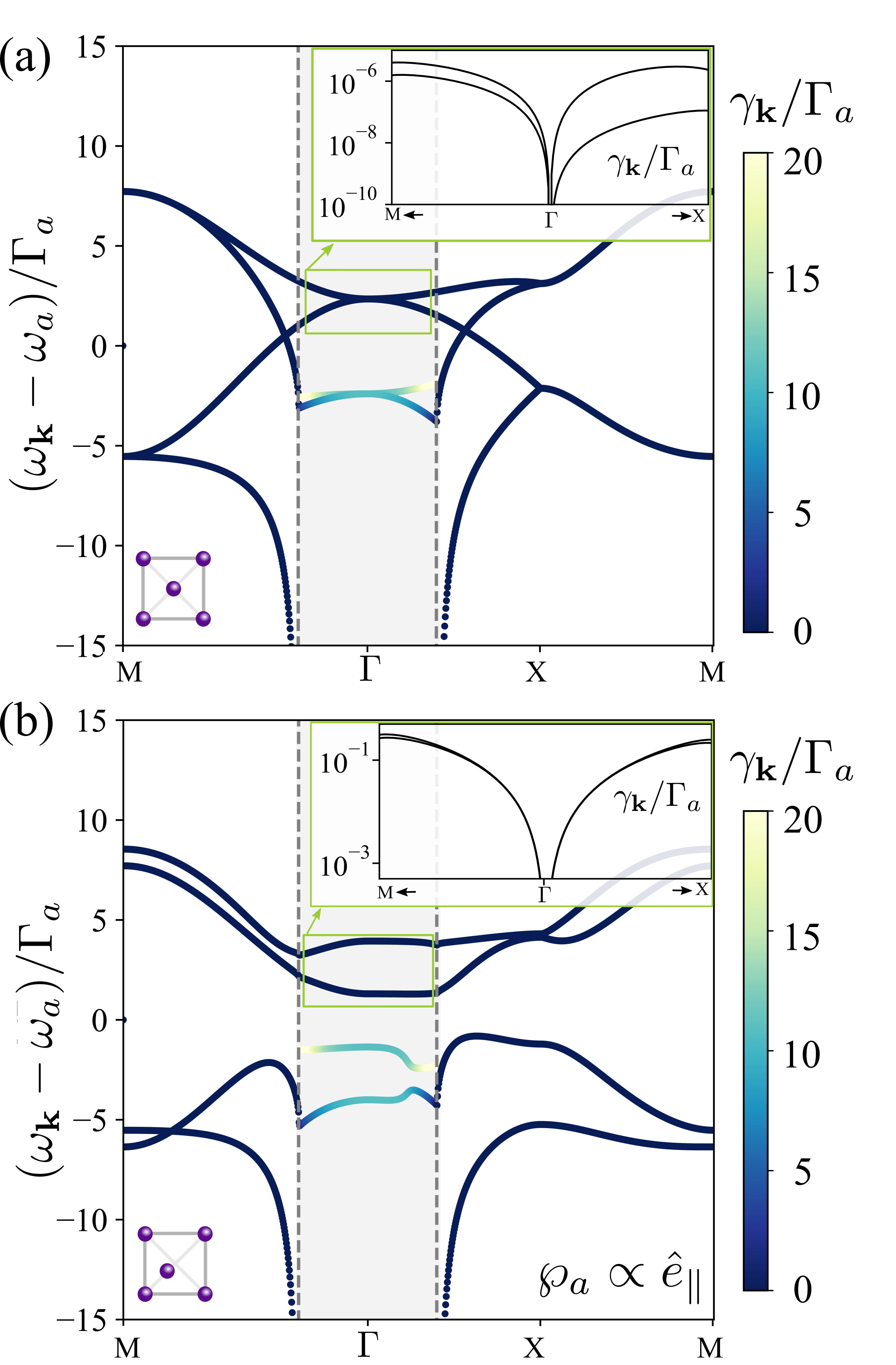}
  \caption{Band structure for in-plane modes of two lattices of identical emitters, $\omega_{1,2} = \omega_a$.  The mode decay rate is shown in color (legends on the right hand side). A detail of the decay rates of the subradiant modes in the continuum range is shown in the insets. (a) Lattice with $\mathbf{T}=a(0.5,0.5)$, with the subradiant modes mapped at $\Gamma$. (b) Lattice with  $\mathbf{T}=a(0.4,0.4)$, with BICs at $\Gamma$. In both panels  $a=0.2\lambda_a$. The light line is marked with dashed lines, and the radiation continuum is shaded in gray. The insets show the decay rate of the two subradiant eigenmodes.}
  \label{figBands}
\end{figure}

The band structure of the in-plane modes of a bipartite square lattice of identical emitters ($\omega_{1,2}=\omega_a$, $\Gamma_{1,2}=\Gamma_a$) 
with periodicity {$a=0.2\lambda_a$} is shown in Fig.~\ref{figBands}. The basis vector of the lattice considered in panel (a) is $\mathbf{T}=(0.5,0.5)a$, such that the array is a Bravais lattice with smaller rotated unit cell, and the four normal modes are degenerate by pairs at normal incidence and at the edges of the Brillouin zone. The plot shows the eigen-energies, $\omega_\mathbf{k} = \omega_a + \Delta_\lambda(\kk_\|) $, where $\Delta_\lambda $ are the eigenvalues of, 
\begin{equation}\label{Eq:delta}
    \hat{\Delta}(\kk_\|) = 
    \begin{bmatrix}
        \hat{\Delta}^{11}(\kk_\|) & \hat{\Delta}^{12}(\kk_\|) \\
        \hat{\Delta}^{21}(\kk_\|) & \hat{\Delta}^{11}(\kk_\|)
    \end{bmatrix}, \, 
\end{equation}
where the $ \hat{\Delta}^{\mu\nu}$ matrices now correspond to the $2\times2$ in-plane blocks. The decay rates, $\gamma_\kk = \Gamma_a + \Gamma_\lambda(\kk_\|)$, with $\Gamma_\lambda$ the eigenvalues of the decay rate matrix of the lattice, are color coded in the band structure. As can be observed in the figure, there are two sets of radiative modes, that is, within the light cone (gray area). However, by looking at their decay rate we see that the two at lower energies are super-radiant, with $\gamma_\kk \gg \Gamma_a$, while the two at higher energies are sub-radiant, with $\gamma_\kk \ll \Gamma_a$. The decay rate of these sub-radiant modes is plotted as an inset, where we see how $\gamma_\kk \ll \Gamma_a$ and $\gamma_\kk \rightarrow 0$ as $\kk_{||}\rightarrow 0$. This is in agreement with the fact that these modes lie beyond the light line of the lattice with smaller unit cell, and are mapped into the zone center by folding. In real space, they correspond to the two dipoles in the unit cell being anti-aligned, a configuration that for symmetry reasons cannot radiate to the far field. In panel (b), a non-Bravais lattice is considered by taking $\mathbf{T}=(0.4,0.4)a$, which lifts the band degeneracies of panel (a). In this case, modes above the light line are actually within the radiation continuum. However, the decay rate of the two subradiant modes at $\kk_{||} = 0$ is  $\gamma_\kk =0 $, as can be seen from symmetry analysis of the eigenstates. 
Normal incidence implies $\hat{\mathbf{S}}^{21}(\kk_{||} = 0)= \hat{\mathbf{S}}^{12}(\kk_{||} = 0)$, and due to the lattice symmetry $ S^{11}_{xy/yx}=0 $ and $S^{12}_{xy/yx} = 0$. Therefore, at the $\mathbf{\Gamma}$ point, Eq.~\ref{Eq:delta} has two pairs of non-degenerate eigenvalues,
\begin{equation} \label{eq:coopshifLatticesum2}
    \Delta_{\pm,i} (\mathbf{k}_{||}=0) =  -\frac{3}{2}\Gamma_a\lambda_a\text{Re}[S^{11}_{ii}\pm S^{12}_{ii}], 
\end{equation}
with decay rates,
\begin{equation} \label{Eq:coopdecayLatticesum2}
    \Gamma_{\pm,i} (\mathbf{k}_{||}=0) =  3\Gamma_a\lambda_a\text{Im}[S^{11}_{ii}\pm S^{12}_{ii}],
\end{equation}
corresponding to one symmetric and one anti-symmetric mode per spatial degree of freedom ($i=\{x,y\}$).
Making use of the fact that $\text{Im}[S^{11}_{xx}] =  \text{Im}[S^{12}_{xx}]- (3\lambda_a)^{-1}$, we find
$  \Gamma_{-} (\mathbf{k}_{||} = 0) = - \Gamma_a$ and $\gamma_\kk=0$ for the subradiant modes. Thus, even if the radiative decay of the antisymmetric modes is larger than in the case of panel (a), it is completely suppressed at normal incidence, as shown in the inset panel. These are then subradiant modes embbeded in the continuum, confirming the existence of BICs at normal incidence. 

Next we study the optical properties of the emitter array by looking at the reflectivity, which can be calculated from the total dipole moment of the lattice excited by an incident plane wave with $s$ or $p$ polarisation (see S.M.), as
\begin{equation}\label{eq:reflectivity}
    R_{\sigma\sigma'}(\mathbf{k}_{\|}) =   \frac{|\mathbf{e}_\sigma \cdot \hat{\boldsymbol{\alpha}}_\text{eff}\cdot\mathbf{e}_{\sigma'}|^2 }{(A\lambda\varepsilon_0/\pi)^2}.
\end{equation}
Here, $A$ is the unit cell area, $\sigma=s,p$, and $\hat{\boldsymbol{\alpha}}_\text{eff}$  is the effective polarisability tensor~\cite{AguiePRL,Baur18,Kolkowski19,Becerril20}. This is a 3$\times$3 matrix that represents the total response of the array in a Cartesian basis and is obtained by summing over the sublattice responses, 
\begin{align}
   \left[ \hat{\boldsymbol{\alpha}}^{-1}_\text{eff}(\omega,\mathbf{k}_{\|})  \right]= \sum_\mu^M\sum_\nu^M \hat{\boldsymbol{{\beta}}}^{\mu\nu}_\text{eff}. 
\end{align}
Here we have introduced a generalised effective polarisability tensor, 
\begin{eqnarray} \label{eq:generalizedeffalpha}
    \left[ \hat{{\boldsymbol{\beta}}}^{-1}_\text{eff}(\omega,\mathbf{k}_{\|})  \right]^{\mu\nu} =   \frac{\mathbbm{1}}{\alpha^\mu(\omega)} \delta_{\mu,\nu}  -  \frac{k^2}{\varepsilon_0}\hat{\mathbf{S}}^{\mu\nu}(\mathbf{k}_{||}),
\end{eqnarray}
which is a $3M\times3M$ tensor (see S.M.).
\begin{figure}[ht]
  \centering
  \includegraphics[width=0.99\columnwidth]{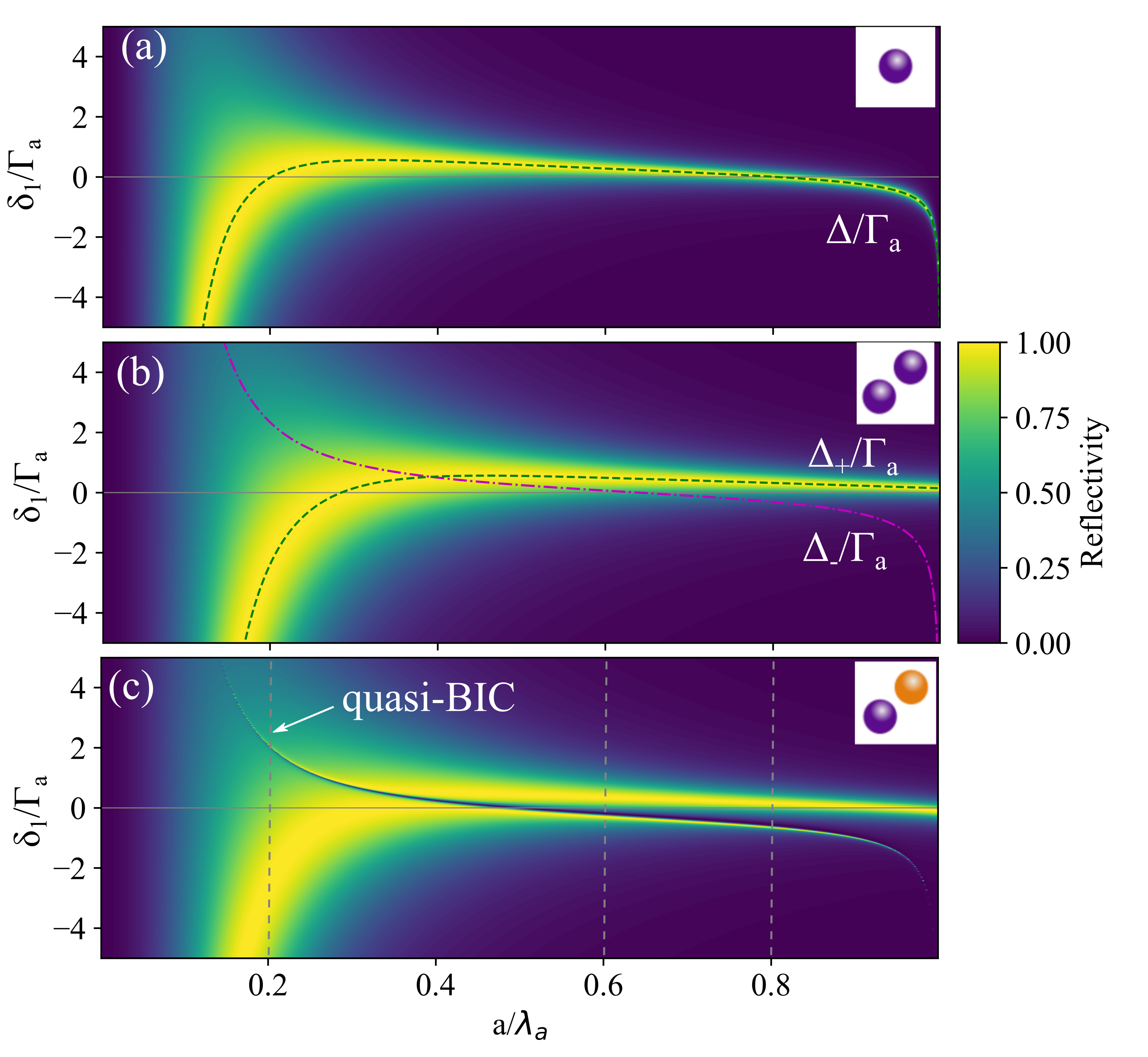}
  \caption{Normal incidence reflectivity as function of detuning and periodicity for a lattice with a basis of one emitter per cell (a), with two identical emitters per cell (b), and with two detuned emitters per cell (c). In (b), two subradiant degenerate modes are marked with a purple dashed-dotted line ($\delta_1 = \Delta_- $). The superradiant mode is marked with a green dashed line ($\delta_1 = \Delta_+ $). (c) Symmetry breaking through detuning one of the emitters ($\omega_2 = \omega_1 + \Gamma_a /2$) allows for the emergence of quasi-BICs.  In (b,c) $\textbf{T} = (a/2,a/2)$.  }
  \label{figReflectivity}
\end{figure}

Figure \ref{figReflectivity} shows the array reflectivity at normal incidence for a square lattice with one (a) and two emitters (b,c) per unit cell, as function of frequency detuning with respect to the emitter's resonance $\delta_1=\omega-\omega_a$ and lattice period, $a/\lambda_a$.  The second emitter is placed at a high symmetry point (the center of the cell), as considered in Fig. 2(a).
Here, a p-polarised incident wave ($\mathbf{E}_0=E_0\hat{\mathbf{u}}_x$) is assumed. Due to lattice symmetry ($\hat{\mathbf{S}}^{\mu\nu}_{xx} = \hat{\mathbf{S}}^{\mu\nu}_{yy}$), the response is equal for both polarizations $R_{ss}=R_{pp}=R$.

Panel (a) corresponds to a lattice with basis of one emitter per unit cell. Through the cooperative shift, $ \Delta (\mathbf{k}_{||}=0) =  -\frac{3}{2}\Gamma_a\lambda_a\text{Re}[S_{xx}(\mathbf{k}_{||}=0)]$, the cooperative response of the lattice gives rise to a resonance for $\delta_1=\Delta$ (green dashed line), due to the collective excitation of the in-plane modes by the incident field ~\cite{shahmoon2017cooperative,alaeeQuantum2020}.
 For shorter periods, $a\ll\lambda_a$, $ \text{Re}[S_{xx}]\rightarrow \infty$ and  the resonance is largely redshifted towards negative detunings. As the periodicity increases, the cooperative shift turns positive and the resonance crosses the zero detuning line, going back to negative detunings as $a\rightarrow\lambda_a$ since $\text{Re}[S_{xx}]\rightarrow \infty$ again. On the other hand, the radiative width decreases as the period increases, giving rise to resonances with very narrow linewidth at $a\rightarrow\lambda_a$~\cite{FJRMP,AuguiePRL,KravetsPRL,ChuAPL,VecchiPRB,AdatoOpt,Wang18,Kravets18,acsnanosupersub}.

  \begin{figure*}[t!]
  \centering
  \includegraphics[width=\textwidth]{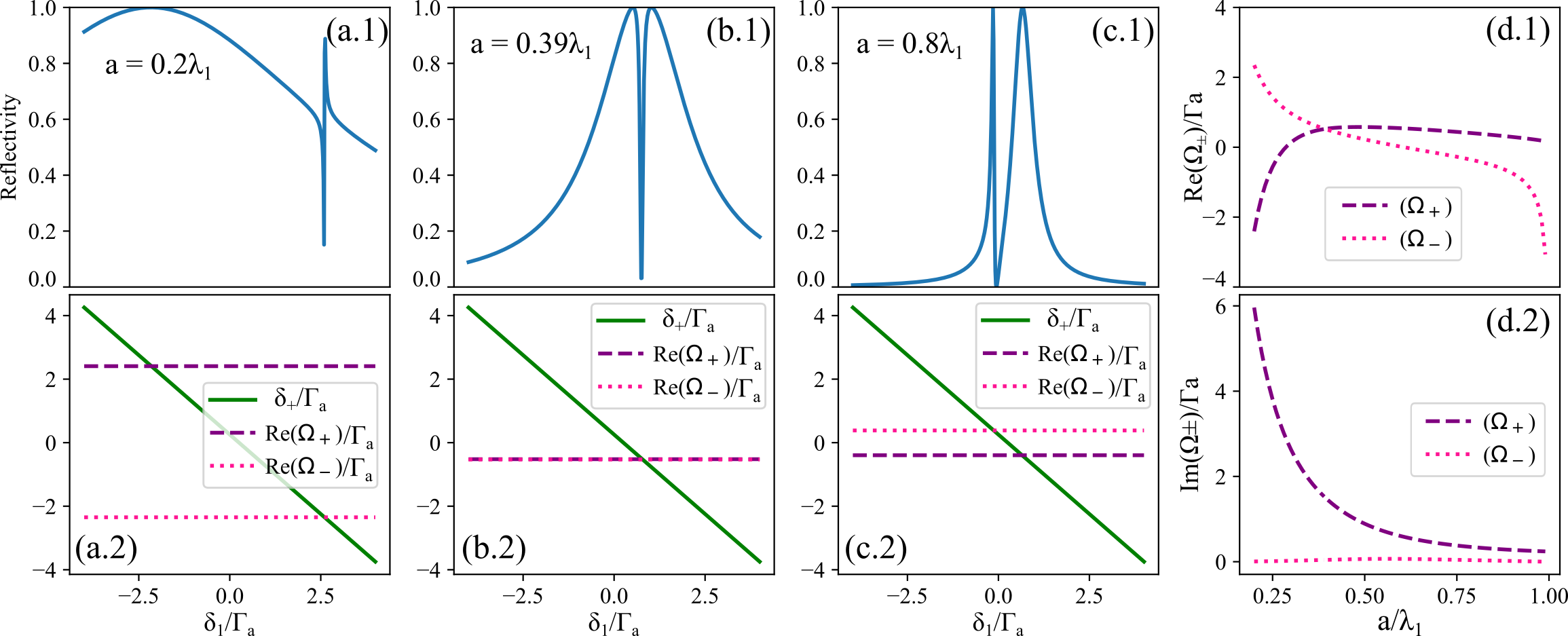}
  \caption{Spectra of detuned bipartite arrays in different parameter regimes. (a.1-c.1) Reflectivity spectrum at normal incidence for lattices with periods $a/\lambda_{a,1}=0.2\,,0.39\,,0.8$ as a function of detuning $\delta_1 = \omega-\omega_1$, with $\omega_2=\omega_1+\Gamma_a/2$. Asymmetric Fano resonances (a,c) and EIT (b) emerge due to the quasi-BICs. The bottom panels (a.2-c.2) show the peak positions as the crossings $\delta_+=\textrm{Re}[\Omega_\pm]$, for the superradiant and subradiant modes in the spectra shown in the three panels above (see Eq.~\ref{Eq:genbetaeff}). Panels (d.1) and (d.2) show real and imaginary parts of $\Omega_\pm$ as function of lattice period.}
  \label{figFano}
\end{figure*}
 
By adding a second emitter per unit cell, the collective response of the lattice is modified through the interplay between the intra- and inter-cell lattice sums, $S^{11}_{xx}$ and $S^{12}_{xx}$, see Eq.~\ref{eq:coopshifLatticesum2}. The reflectivity spectrum of the same system as in Fig.~\ref{figBands}(a) is shown in Fig.~\ref{figReflectivity}~(b). The superradiant mode, $\delta_1=\Delta_{+}$, is marked with a dashed green line, and the subradiant mode, $\delta_1=\Delta_{-}$, with a dashed-dotted purple line. 
The interplay between both sublattices modifies the behaviour of the resonance reflectivity peak compared to that of panel (a). 
First, for $a\ll\lambda_a$, the lattice sums $\textrm{Re}[S^{12}_{xx}]\rightarrow 2 \textrm{Re}[S^{11}_{xx}]$, such that the cooperative shift of the superradiant mode $\Delta_{+}\rightarrow 2\Delta$, while that of the subradiant modes $\Delta_{-}\rightarrow -\Delta$. Hence, while the superradiant mode lies at negative detuning for low periods, the subradiant one lies at positive detuning. Next, as the periodicity increases, there is a value of the period where $\textrm{Re}\left[ S^{12}_{xx}\right]=0$, implying that the four in-plane eigenmodes are degenerate, and the superradiant and subradiant modes cross. For the chosen lattice, this occurs at $a=0.39\lambda_a$ (see 
Fig. S3 in the S.M.). Thus, for larger values of the period the subradiant mode is the one lying at lower detuning, while the superradiant one stays at higher detuning. In fact, when $a\rightarrow\lambda_a$, $\textrm{Re}[S^{12}_{xx}]\rightarrow - \textrm{Re}[S^{11}_{xx}]$, resulting in $\Delta_+\rightarrow0$, and the superradiant mode tends to zero detuning, in sharp contrast to panel (a). 

Next, we introduce a small perturbation in the system which breaks its rotational symmetry and allows us to access the subradiant modes from the far field by transforming them into quasi-BICs. We do this by detuning one of the emitters in the unit cell. This strategy has been used to individually address lattice dark states, showing how they can store and release single photons \cite{rubies-bigordaPhoton2022,ballantineQuantum2021,jenCooperative2016}. The optical response of this type of lattice is shown in Fig.~\ref{figReflectivity}~(c), for $\omega_2 = \omega_1 + \Gamma_a/2$.  The symmetry breaking results in a finite linewidth for the subradiant mode, which thus transforms into a quasi-BIC that can couple to the far field and therefore, becomes visible at normal incidence. Interestingly, their radiative width is small, $\Gamma_- \ll \Gamma_a$, and controllable through the system's parameters.



%
%
We now study in detail the optical properties of the bipartite lattice with broken sublattice symmetry.  
In order to explore the contribution of the superradiant and subradiant modes to the collective response, we  analytically recast the effective polarizability as the sum of two contributions, corresponding to the superradiant (+) and subradiant or quasi-BIC (-) modes (see S.M. and Ref. \cite{acsnanosupersub} for a related expression in the context of plasmonic lattices), 
\begin{equation}\label{Eq:genbetaeff}
      \mathbf{\alpha}^{xx}_\textrm{eff} (\omega, \mathbf{k_{||}}) =  \alpha_0\Gamma_a \left[ \frac{ \Sigma_+ (\mathbf{k_{||}}) }{ \delta_+ (\omega) - \Omega_+ (\mathbf{k_{||}}) }  + \frac{ \Sigma_- (\mathbf{k_{||}})}{ \delta_+ (\omega) - \Omega_- (\mathbf{k_{||}}  ) } \right], 
\end{equation}
where $\alpha_0 = 6\pi\epsilon_0c^3/\omega_a^3 $ and the decay rate of the emitters is assumed to be $\Gamma_1 = \Gamma_2 = \Gamma_a$. The terms $\Sigma_\pm$ and $\Omega_\pm$ are defined respectively as, $\Sigma_\pm(\mathbf{k_{||}})=   \frac{1}{2}  \pm  {S}^{12}_{xx}(\mathbf{k_{||}}) \left[\left({S}^{12}_{xx} (\mathbf{k_{||}})\right)^2 + (\delta_-)^2 \right]^{-1/2} 
$ and 
$\Omega_\pm (\mathbf{k_{||}})=  {S}^{11}_{xx} (\mathbf{k_{||}})\pm \textrm{sign} \left( -\Delta_{xx}^{12} \right) \sqrt{\left({S}^{12}_{xx}(\mathbf{k_{||}})\right)^2 + (\delta_-)^2 } +i\Gamma_a/2$, both expressions being solely functions of the incident field's momentum. The term $\delta_- = (\omega_1-\omega_2)/2$ is a constant and the frequency dependency is solely introduced in Eq.~\ref{Eq:genbetaeff} through the term $\delta_+ (\omega)= -(\delta_1(\omega)+\delta_2(\omega))/2$. Given that the reflectivity of the lattice is proportional to $|\hat{\mathbf{\alpha}}_{eff}|^{2}$ (see Eq.~\ref{eq:reflectivity}) the interference between the two modes gives rise to a very rich spectrum.

Figure \ref{figFano}(a-c) shows the reflectivity spectrum of the array at fixed values of the lattice period $a/\lambda_{a,1}=0.2\,,0.39\,,0.8$, marked with vertical lines
in Fig.~\ref{figReflectivity}~(c).
These spectra show three different regimes of interaction, defined by the relative position between the superradiant and the subradiant (quasi-BIC) modes, and their radiative widths. 
In the lower row, panels (a.2-c.2) display the corresponding resonance frequency of the superradiant modes as the crossing between $\delta_+$ (green line) and the real part of $\Omega_+$ (dashed purple), as given by Eq.~\ref{Eq:genbetaeff}. Correspondingly, the quasi-BIC is marked by the crossing  between $\delta_+$ and $\Omega_-$ (dotted pink). 
As seen in panels (d.1) and (d.2), at short periods the superradiant mode lies at lower frequencies than the quasi-BIC ($\textrm{Re}(\Omega)_+<\textrm{Re}(\Omega)_-$), and has a very large radiative width ($\textrm{Im}(\Omega)_+\gg \Gamma_a $). As a consequence, the quasi-BIC mode emerges as a very sharp and asymmetric spectral feature (see panel a, for $a=0.2\lambda_a$ ). 
As the period increases the two modes approach in frequency and become degenerate when $\text{Re}[S_{xx}^{12}]=0$, in this case at $a=0.39\lambda_a$. This results in an electromagnetically induced transparency window \cite{RevModPhys.77.633,PhysRevLett.117.243601,PhysRevLettJenkins}, as shown in panel (b). The coherent interaction between the degenerate modes leads to a symmetric spectrum with a narrow window of complete transparency. Finally, for longer periods the superradiant mode lies at higher frequencies  ($\textrm{Re}(\Omega)_+>\textrm{Re}(\Omega)_-$, see d.1), and its radiative width decreases (see panel d.2). This results in a narrow and asymmetric quasi-BIC peak, separated by a zero reflectivity point from a broader peak (see panel c, $a=0.8\lambda_a$), a qualitatively different different Fano profile from that shown in panel a. These results show how the reflectivity spectrum of the emitter array can be tuned by geometrical means, such as lattice periodicity.  

\begin{figure}[t]
  \centering
  \includegraphics[width=\columnwidth]{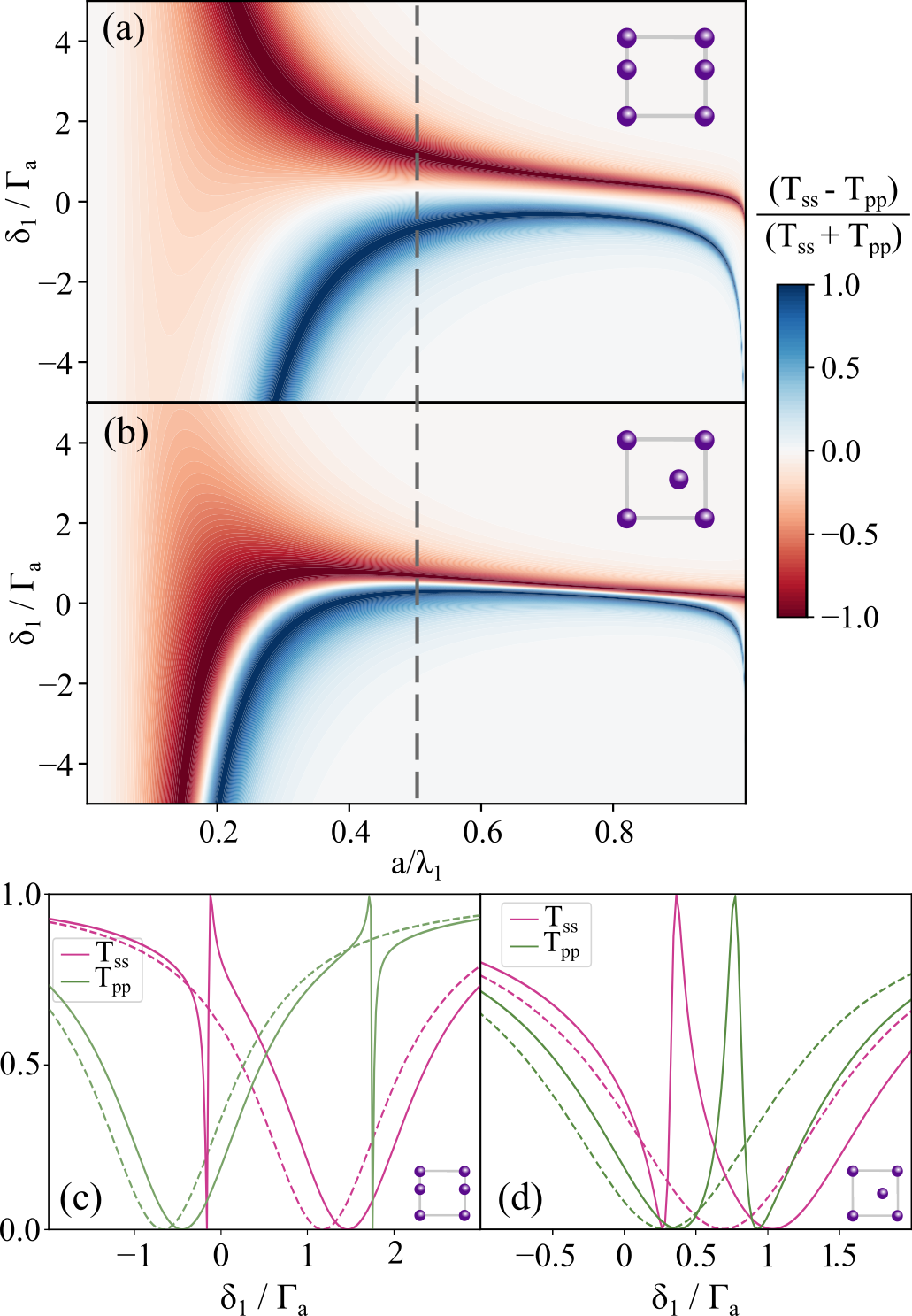}
  \caption{Visibility of transmission at normal incidence for a lattice with basis vector $\mathbf{T}=(0,0.6)a$ (a), and $\mathbf{T}=(0.7,0.5)a$ (b) as function of the lattice period $a/\lambda_a$ and frequency detuning $\delta_1/\Gamma_a$. The grey dashed line indicates the lattice period $a/\lambda_1=0.5$ fixed in panels (c-d), for the basis vectors of (a-b) respectively. Panels (c-d) show the transmission of each polarization, $T_{ss}$ in pink and $T_{pp}$ in green, as function of the detuning $\delta_1/\Gamma_a$. The dashed lines correspond to lattices with $\omega_2=\omega_1$ while the solid line is used for lattices with detuned emitters, $\omega_2=\omega_1 + \Gamma_a/2$.}
  \label{figtrans}
\end{figure}

As we have shown, the optical properties of the lattice are strongly determined by the interplay between the two sublattices through cooperative dipole exchanges between emitters. As a consequence, the array's reflectivity (or transmission), resonance frequency and bandwidth can be controlled and linear optical elements can be designed~\cite{wangDesign2017,arxiv.Genes}. We now exploit this to show how the bipartite emitter arrays can serve as efficient polarizers and identify different regimes of operation.  
In order to design emitter arrays that act as polarizers, we consider geometrical configurations that do not have a diagonal mirror plane as the one studied above. This results in
$S^{\mu\nu}_{xx} \neq S^{\mu\nu}_{yy}$
and as a consequence the response to $s-$ and $p-$polarized incident field is different, allowing to 
filter the polarization of the transmitted light.
To asses the efficiency of the biparte emitter lattices as polarizer we plot the visibility of the transmitted components, $(T_{ss}-T_{pp}) / (T_{ss}+T_{pp})$. 
This measure of the polarising efficiency 
is depicted in Fig.~\ref{figtrans} as function of the lattice period $a/\lambda_{a1}$ and frequency detuning for two representative lattices. Panel (a) corresponds to the lattice with emitters aligned along the $y-$axis (basis vector $\mathbf{T}=(0,0.6)a$), which reduces to a rectangular lattice \cite{wangDesign2017}, 
The large $x/y$ asymmetry of inter-sublattice interactions gives rise to a highly efficient polariser, with the visibility displaying a broad peak at positive detuning reaching values of $+1$. At these points only $s-$polarised waves are transmitted, and transmission of $p-$polarised light is completely suppressed (transmission plots can be found in the S.M.). Conversely, at negative detunings visibility of $-1$ is reached, with only $p-$polarised light being transmitted. This behaviour is explained by the fact that Re[$S^{12}_{yy}$]>0 and Re[$S^{12}_{xx}$]<0 for all periods (see S.M.). 
In contrast, for the configuration shown in panel (b), with basis vector $\mathbf{T}=(0.7,0.5)d$, 
the high visibility peak of $p-$polarized transmission shifts to negative detunings.
Next, we exploit the quasi-BIC resonances generated with detuned emitters to achieve control of the polarizer bandwidth. Panels (c-d) plot the transmittance spectra for the two lattices, with periodicity $a/\lambda_a = 0.5$, indicated by a vertical dashed line in the contourplots. $T_{ss}$ is shown in red and $T_{pp}$ in blue.   
Two cases are shown in each panel: the spectra for identical emitters in both sublattices ($\omega_2=\omega_1$, dashed lines), and with detuned emitters ($\omega_2=\omega_1 + \Gamma_a/2$, solid lines). Complete contour plots for detuned emitters can be found in the S.M.
In both cases, the presence of the quasi-BIC modes allows us to realise narrow windows of efficient polarising effect.  
Note it is possible to tune perfect transmittance for one polarization while completely cancelling the other for very narrow frequency windows, as shown for instance in panel (d) at detuning $\delta_1\sim0.5\Gamma_a$, with $T_{ss}=1$ and $T_{pp}=0$. 
In conclusion, we have proposed the use of two-dimensional non-Bravais lattices of quantum emitters to realize subradiant modes that are not spatially confined to the array. Perfectly periodic emitter arrays with more than one sublattice support BICs, which are completely dark (subradiant) modes lying within the radiation continuum and thus accessible from the far field. Their frequency can be tuned through the geometrical parameters of the lattice. Symmetry breaking through, e.g. sublattices of slightly detuned emitters, results in subradiant modes of finite radiative width but high quality factor which emerge as sharp Fano peaks in the optical spectrum. By tuning the lattice geometry, it is possible to achieve an electromagnetically induced transparency window. The low radiative width of the subradiant modes studied here can be potentially exploited for the purpose of quantum information storage. Two-dimensional lattices of quantum emitters can be experimentally realized in a cold atoms platform \cite{ruiSubradiant2020,srakaew2022subwavelength}. 
Furthermore, we show how to employ these lattices as quantum metasurfaces acting as polarizers, with controllable resonance frequency and bandwidth.
Finally, symmetry-protected BICs posses a topological charge \cite{Zhen14} which can be imprinted on the incident light in order to create non-trivial beams. 


\begin{acknowledgements}
The authors acknowledge fruitful discussions with Alejandro Gonz\'alez Tudela.
We acknowledge funding from: Funda\c c\~ao para a Ci\^encia e a Tecnologia and Instituto de Telecomunica\c c\~oes under projects UTAP-EXPL/NPN/0022/2021 and UIDB/50008/2020, the FCT-CEEC Individual program (CEECIND/02947/2020), the Spanish Ministry for Science and Innovation (Grant No. RYC2021-031568-I), and the CAM (Y2020/TCS-6545). 
M.B.d.P. acknowledges support from the Spanish MICINN (PID2019-109905GA-C2) and from the IKUR Strategy under the collaboration agreement between Ikerbasque Foundation and DIPC on behalf of the Department of Education of the Basque Government, as well the Basque Government Elkartek program (KK-2021/00082) and the Centros Severo Ochoa AEI/CEX2018-000867-S from the MICINN.

\end{acknowledgements}

\bibliographystyle{apsrev4-1}
\bibliography{references}

\end{document}


\title{Supplemental Material:
\\Bound states in the continuum in subwavelength emitter arrays}

\author{Mar\'ia Blanco de Paz}
\affiliation{Donostia International Physics Center, 20018 Donostia-San Sebasti\'an, Spain}
\affiliation{Instituto de Telecomunica\c c\~oes, Instituto Superior Tecnico-University of Lisbon, Avenida Rovisco Pais 1, Lisboa, 1049-001 Portugal}

\author{Paloma A. Huidobro}
\email{p.arroyo-huidobro@uam.es}
\affiliation{Instituto de Telecomunica\c c\~oes, Instituto Superior Tecnico-University of Lisbon, Avenida Rovisco Pais 1, Lisboa, 1049-001 Portugal}
\affiliation{Departamento de F\'{i}sica Te\'orica de la Materia Condensada and Condensed Matter Physics Center (IFIMAC), Universidad Aut\'onoma de Madrid, E-28049 Madrid, Spain}
\maketitle

\section{Classical scattering problem}

We consider a two-dimensional (2D) lattice of quantum emitters in a non-Bravais lattice. The emitters positions are given by $\mathbf{R}_{n}^{\mu} = \mathbf{T}_\mu + \mathbf{R}_n $, with $\{\mathbf{R}_n\}$ giving the positions of all unit cells ($n=1,\cdots, N$) and $\{\mathbf{T}_\mu\}$ being the basis vectors of the sublattices ($\mu=1,\cdots, M$), which join the origin of the unit cell with each of the emitters in the basis. In a two-level system approximation, the optical response of each emitter is characterised by a polarizability, 
\begin{equation} \label{eq:polariz}
    \alpha (\omega)=-\alpha_{0}\frac{ \Gamma_{a} /2} {\delta + i \Gamma_{a}/2} , 
\end{equation}
where $ \delta= \omega-\omega_{a}$ is the frequency detuning with respect to the resonance frequency of the two level system, $\omega_a$, $\Gamma_a$ its decay rate, and $ \alpha_{0} = 6 \pi \varepsilon_{0}/ k_{a}^{3}$, with $k_{a}=\omega_{a}/c$. Since the size of the emitters is small enough with respect to the relevant wavelength of the electromagnetic field we can apply the dipole approximation and write a self-consistent equation for the dipole moments of the emitters, 
\begin{equation} 
    \frac{1}{\alpha^\mu(\omega)}\mathbf{p}_{n}^\mu =  \mathbf{E}_{\mathrm{inc}}(\mathbf{R}_{n}^\mu) + 
    \sideset{}{'}\sum_{m=1}^{N} \sum_{\nu=1}^M
     \frac{k^{2}}{\varepsilon_{0}} \hat{\mathbf{G}}(\mathbf{R}_{n}^\mu,\mathbf{R}_{m}^\nu)\mathbf{p}_{m}^\nu\,.
\end{equation}
Here, the primed sum indicates that terms $m=n$ are excluded from the sum when $\mu=\nu$ to exclude self-interactions, and we take equal emitters for each sublattice, but we allow for the polarisability of the emitters in different sublattices to be different. The photon-mediated interactions between the emitters are given by the free-space Green's dyadic,
\begin{eqnarray}
    &&\hat{\mathbf{G}}(\mathbf{r},\mathbf{r}') = \frac{1}{4\pi}\left(  \mathbbm{1} + \frac{1}{k^2} \nabla \otimes \nabla \right) \frac{e^{ik | \mathbf{r}-\mathbf{r}'|}}{| \mathbf{r}-\mathbf{r}'|},
\end{eqnarray}
where $k=\omega/c$.

For a given incident field, $\mathbf{E}_{\mathrm{inc}}(\mathbf{k}_{\|}) = \mathbf{E}_0 e^{i \mathbf{k} \cdot \mathbf{r} } $, with $\mathbf{k}=(\mathbf{k}_{\|},k_z) $, $\mathbf{k}_{||} $ the momentum component in the plane of the array and $k_z=\sqrt{k^2-|\mathbf{k}_{\|}|^2}$ the momentum component perpendicular to the array, we apply Bloch's theorem, $\mathbf{p}_n^\mu = \mathbf{p}^\mu e^{i\mathbf{k}_{||} \mathbf{R}_n} $, and arrive at a matrix equation for the dipole moments, 
\begin{equation} \label{eq:coupleddipoles}
    \sum_{\nu=0}^M \left( \frac{1}{\alpha^\mu(\omega)} \delta_{\mu,\nu}  - \frac{k^{2}}{\varepsilon_{0}}  \hat{\mathbf{S}}^{\mu\nu}(\omega,\mathbf{k}_{||}) \right) \mathbf{p}^\mu = \mathbf{E}_{\mathrm{inc}}^\mu\,.
\end{equation}
Here,  $\mathbf{E}_{\mathrm{inc}}^\mu=\mathbf{E}_0 e^{i\mathbf{k}_{||} \mathbf{T}_\mu}$, and $\hat{\mathbf{S}}^{\mu\nu}(\omega,\mathbf{k}_{||})$ stands for the Fourier transform of the dipole-dipole interaction. This is the lattice sum, 
\begin{equation} \label{eq:latticesums}
    \hat{\mathbf{S}}^{\mu\nu}(\omega,\mathbf{k}_{||}) =  \sideset{}{'}\sum_{m=1}^{N}  \hat{\mathbf{G}}(\mathbf{R}_{m} + \mathbf{T}^\nu -\mathbf{T}^\mu) e^{-i\mathbf{k}_{||} \mathbf{R}_{m}}\,
\end{equation}
where the primed symbol indicates that the sum excludes the term $m=1$ when $\mu=\nu$, such that we do not include self-interactions. In compact matrix form, Eq. \ref{eq:coupleddipoles} reads as, 
\begin{equation} \label{eq:coupleddipolesM}
    \hat{\textbf{M}}\, \mathbf{p} = \mathbf{E}_{\mathrm{inc}}\,,
\end{equation}
with $ \hat{\textbf{M}}^{\mu\nu}(\omega,\mathbf{k}_{||}) = 
    \delta_{\mu,\nu}/\alpha^\mu(\omega)  - (k^{2}/\varepsilon_{0})  \hat{\mathbf{S}}^{\mu\nu}(\omega,\mathbf{k}_{||}) $. 
Through the lattice sums, which depend only on the geometry of the lattice and not on the properties of the emitters, the above matrix equation models the collective electromagnetic interaction between all the quantum emitters within the single excitation approximation. 

\section{Quantum formalism}
Within the dipole approximation, the Hamiltonian of a collection of quantum emitters interacting with the quantized electromagnetic field is, 
\begin{align}
    H = \hbar\sum_{j=1}^{N_A}\omega_j \sigma_{ee}^j+ \hbar \sum_{\sigma= s,p} \int \textrm{d}^3 k \,\,\,\omega_k a_{\mathbf{k}\sigma}^\dagger  a_{\mathbf{k}\sigma} - 
    \sum_{j=1}^{N_A} \mathbf{d}_j \cdot \mathbf{E}(\mathbf{R}_j)
\label{eq:H}
\end{align}
Here, $N_A$ is the total number of atoms in the array, with dipole moments $d_j$, transition frequencies $\omega_j$ and free space decay rates $\Gamma_j$. The sum runs over all emitters and the dipole operators are vector quantities representing the three space directions. The third term in the Hamiltonian represents the dipolar interaction between the emitters and electromagnetic field at the position of the emitter. The field is given by the input field and the field scattered by all the other emitters, 
\begin{equation}
    \mathbf{E}(\mathbf{r},\omega) =   \mathbf{E}_\textrm{inc}(\mathbf{r},\omega) + \frac{\omega^2}{\epsilon_0 c^2} \sum_{i=1}^{N_A} \mathbf{G} (\mathbf{r},\mathbf{r}_i,\omega) \mathbf{d}_i(\omega).  
\end{equation}
Since the propagation of both classical and quantum fields is governed by Maxwell's equations, the above equation is valid also for the quantized field operator, $\hat{\mathbf{E}}(\mathbf{r}) $, and dipole operator, $\hat{\mathbf{d}}(\mathbf{r}) $. 

Then, we take the Markov approximation. This assumes we can take we can take $\omega\approx\omega_i$, the resonance frequency of the atoms, thus neglecting frequency dispersion in the Green's function. This is a good approximation since the atomic optical response has a very narrow bandwidth around its resonance frequency. With this, 
\begin{equation}
    \hat{\mathbf{E}}(\mathbf{r}) =   \hat{\mathbf{E}}_\textrm{inc}(\mathbf{r}) +  \sum_{i=1}^{N_A} \frac{\omega_i^2}{\epsilon_0 c^2} \mathbf{G} (\mathbf{r},\mathbf{r}_i,\omega_i)  \hat{\mathbf{d}}_i,  
\end{equation}
such that the interaction term is 
\begin{equation}
    H = -  
    \sum_{j=1}^{N_A}  \sum_{i\neq j} \hat{\mathbf{d}}_j \cdot \frac{\omega_i^2}{\epsilon_0 c^2} \mathbf{G} (\mathbf{R}_j,\mathbf{R}_i,\omega_i) \cdot \hat{\mathbf{d}}_i 
\end{equation}
where the dipole operators are $\hat{\mathbf{d}}_i = \boldsymbol{\wp}_i^* \hat{\sigma}_{eg}^i + \boldsymbol{\wp}_i \hat{\sigma}_{ge}^i $, with $\wp$ the dipole matrix element of the relevant transition. Next, in the Born-Markov approximation, the adiabatic elimination of the reservoir degree of freedom yields a non-Hermitian effective spin Hamiltonian: 
\begin{align}
    H = \hbar \sum_{j=1}^{N_A}\left(\omega_j - i \frac{\Gamma_j}{2}\right)\sigma_{ee}^j-  \sum\limits_{\substack{i=0 \\ i\neq j}}^{N_A}  \frac{\omega_i^2}{\epsilon_0 c^2} |\boldsymbol{\wp}_i||\boldsymbol{\wp}_j| \left[ \hat{\boldsymbol{\wp}}_j^* \cdot \mathbf{G} (\mathbf{R}_j,\mathbf{R}_i,\omega_i) \cdot \hat{\boldsymbol{\wp}}_i \right] \sigma_{eg}^j \sigma_{ge}^i
\label{eq:Heff}
\end{align}
with $\hat{\boldsymbol{\wp}}_i = \boldsymbol{\wp}_i/|\boldsymbol{\wp}_i| $. 

Considering the case of a non-Bravais lattice, we can express the sums as,
\begin{align}
    H = \hbar \sum_{n=1}^{N}\sum_{\mu=1}^{M}\left(\omega_n^\mu - i \frac{\Gamma_n^\mu}{2}\right)\sigma_{ee}^{n,\mu} -  \sideset{}{'}\sum\limits_{\substack{m=1 \\ n=1 }}^{N} \sum\limits_{\substack{\mu=1 \\ \nu=1 }}^{M}    \frac{\left(\omega_m^\nu\right)^2}{\epsilon_0 c^2} |\boldsymbol{\wp}_m^\nu||\boldsymbol{\wp}_n^\mu| \left[ \hat{\boldsymbol{\wp}}_n^{\mu*} \cdot \mathbf{G} (\mathbf{R}_n^\mu,\mathbf{R}_m^\nu,\omega_m^\nu) \cdot \hat{\boldsymbol{\wp}}_m^\nu \right] \sigma_{eg}^{n,\mu} \sigma_{ge}^{m,\nu}
\label{eq:Heffsub00}
\end{align}
where the primed sum indicates that the term $n=m$ is excluded from the sum when $\mu=\nu$. All emitters within the same sublattice are identical, so $\omega_n^\mu = \omega^\mu $ and $\Gamma_n^\mu = \Gamma^\mu$. Additionally, we consider emitters in different sublattices that are only slightly detuned, $(\omega^\mu - \omega^\nu) / \omega^\mu  \ll 1 $, such that in the interaction Hamiltonian we take $\omega^\nu = \omega^{\nu=1}=\omega_a$, with $\omega_a$ being the resonance frequency of atoms in a reference sublattice. Last, we also assume $(\Gamma^\mu - \Gamma^\nu) / \Gamma^\mu  \ll 1 $ and $(\boldsymbol{\wp}^\mu - \boldsymbol{\wp}^\nu) / \boldsymbol{\wp}^\mu  \ll 1 $, such that $ \Gamma^\mu \approx \Gamma_a $  and $\boldsymbol{\wp}^\nu \approx \boldsymbol{\wp}_a $ for all $\mu$.  With this we can write, 
\begin{align}
    H &= \hbar \sum_{n=1}^{N}\sum_{\mu=1}^{M}\left(\omega^\mu - i \frac{\Gamma_a}{2}\right)\sigma_{ee}^{n,\mu} -   \sideset{}{'}\sum\limits_{\substack{m=1 \\ n=1 }}^{N}\sum\limits_{\substack{\mu=1 \\ \nu=1 }}^{M}    \frac{\omega_a^2 |\boldsymbol{\wp}_a|^2 }{\epsilon_0 c^2} \left[ \hat{\boldsymbol{\wp}}^{*} \cdot \mathbf{G} (\mathbf{R}_n^\mu,\mathbf{R}_m^\nu,\omega_a) \cdot \hat{\boldsymbol{\wp}} \right] \sigma_{eg}^{n,\mu} \sigma_{ge}^{m,\nu} \\
    &= \hbar \sum_{n=1}^{N}\sum_{\mu=1}^{M}\left(\omega^\mu - i \frac{\Gamma_a}{2}\right)\sigma_{ee}^{n,\mu} -  \hbar \sideset{}{'}\sum\limits_{\substack{m=1 \\ n=1 }}^{N}\sum\limits_{\substack{\mu=1 \\ \nu=1 }}^{M}    \frac{3\pi\Gamma_a c  }{\omega_a} \left[ \hat{\boldsymbol{\wp}}^{*} \cdot \mathbf{G} (\mathbf{R}_n^\mu,\mathbf{R}_m^\nu,\omega_a) \cdot \hat{\boldsymbol{\wp}} \right] \sigma_{eg}^{n,\mu} \sigma_{ge}^{m,\nu}
\label{eq:Heffsub}
\end{align}
where we have used the fact that the free space decay rate, $\Gamma_a = |\boldsymbol{\wp}_a|^2 \omega_a^3 / (3\pi\epsilon_0\hbar c^3) $.

For an infinite array we consider Bloch modes, 
\begin{align}
 S^\dagger_\mathbf{k}&=\frac{1}{\sqrt{N}}\sum_{n=1}^N\sum_{\mu=1}^M \sigma^{n,\mu}_{eg} e^{i\mathbf{k}\cdot\mathbf{r}_n},  \\
  \sigma^{n,\mu}_{eg} &=\frac{1}{\sqrt{N}}\sum_{\mathbf{k}} S_\mathbf{k} e^{-i\mathbf{k}\cdot\mathbf{r}_n}, 
\end{align}  
where $\mathbf{k}$ is a wavevector in the Brillouin zone, and we have, 
\begin{align}
    H/\hbar &=  \sum_{\mathbf{k}}\sum_{\mu=1}^{M} \sum_{\nu=1}^{M} \left[  \left(\omega^\mu - i \frac{\Gamma_a}{2}\right) \mathbbm{1}\delta_{\mu\nu}  - \frac{3\pi\Gamma_a c  }{\omega_a}    \sideset{}{'}\sum\limits_{m}  \hat{\boldsymbol{\wp}}^{*} \cdot \left[ \mathbf{G} (\mathbf{R}_m + \mathbf{T}^\nu - \mathbf{T}^\mu,\omega_a)  e^{-i\mathbf{k}\cdot \mathbf{R}_m} \right] \cdot \hat{\boldsymbol{\wp}}  \right] S_\mathbf{k}^\dagger S_\mathbf{k}
\end{align}
where the identity in the first terms is the $3\times3$ identity matrix, accounting for the 3 spatial components. 

In the above equation, we identify the lattice sum, 
\begin{equation} \label{eq:latticesums2}
    \hat{\mathbf{S}}^{\mu\nu}(\mathbf{k}_{||},\omega_a) =  \sideset{}{'}\sum_{m}  \hat{\mathbf{G}}(\mathbf{R}_{m} + \mathbf{T}^\nu -\mathbf{T}^\mu,\omega_a) e^{-i\mathbf{k}_{||} \mathbf{R}_{m}}, 
\end{equation}
such that we can write the Hamiltonian as, 
\begin{align}
    H/\hbar &=   \sum_{\mathbf{k}} \sum_{\mu,\nu}^{M}  \left[  \left(\omega^\mu - i \frac{\Gamma_a}{2}\right) \mathbbm{1}\delta_{\mu\nu}  - \frac{3\pi\Gamma_a c  }{\omega_a}    \hat{\boldsymbol{\wp}}^{*} \cdot \hat{\mathbf{S}}^{\mu\nu}(\mathbf{k}_{||},\omega_a) \cdot \hat{\boldsymbol{\wp}}  \right] S_\mathbf{k}^\dagger  S_\mathbf{k} \\
    &= \sum_{\mathbf{k}} \sum_{\mu,\nu}^{M}  \left[  \left(\omega^\mu - i \frac{\Gamma_a}{2}\right) \mathbbm{1}\delta_{\mu\nu}  + \left(\hat{\Delta}^{\mu\nu} (\mathbf{k}) - i \frac{\hat{\Gamma}^{\mu\nu} (\mathbf{k})}{2}\right) \right] S_\mathbf{k}^\dagger  S_\mathbf{k}.
\end{align}
Here we have introduced the cooperative shift and decay tensors, 
\begin{align}
    \hat{\Delta}^{\mu\nu} (\mathbf{k}) &=  -\frac{3\pi\Gamma_a c  }{\omega_a} \textrm{Re}\left[\hat{\boldsymbol{\wp}}^{*} \cdot \hat{\mathbf{S}}^{\mu\nu}(\mathbf{k}_{||},\omega_a) \cdot \hat{\boldsymbol{\wp}}  \right] \\
    \hat{\Gamma}^{\mu\nu} (\mathbf{k}) &=  \frac{6\pi\Gamma_a c  }{\omega_a} \textrm{Im}\left[\hat{\boldsymbol{\wp}}^{*} \cdot \hat{\mathbf{S}}^{\mu\nu}(\mathbf{k}_{||},\omega_a) \cdot \hat{\boldsymbol{\wp}}  \right]. 
\end{align}

Finally, we can write, 
\begin{align}
    H/\hbar &=   \sum_{\mathbf{k}} \sum_{\mu,\nu}^{M}  \mathbf{N}_\mathbf{k}^{\mu\nu} S_\mathbf{k}^\dagger S_\mathbf{k},
\end{align}
with
\begin{align}
    \mathbf{N}_\mathbf{k}^{\mu\nu}(\mathbf{k}_{||},\omega_a) =   \left(\omega^\mu - i \frac{\Gamma_a}{2}\right) \mathbbm{1}\delta_{\mu\nu}  + \left(\hat{\Delta}^{\mu\nu} (\mathbf{k}) - i \frac{\hat{\Gamma}^{\mu\nu} (\mathbf{k})}{2}\right).
\end{align}
The eigenstates satisfy, 
\begin{align}
    H |\Psi_\mathbf{k}\rangle = \hbar \left(\omega_k - i \frac{\gamma_k}{2}\right)  |\Psi_\mathbf{k}\rangle,
\end{align}
so that the real and imaginary part of the eigenvalues of $\mathbf{N} $ give the cooperative frequency shift and the collective decay rate, $\omega_\mathbf{k}$ and $\gamma_\mathbf{k}$. Clearly, we see that the eigenstates of the system found using a quantum formalism are the same as those found using the classical approach, provided that the intensity is low and the transitions are not saturated. Thus, once the eigenfrequencies and eigenstates are found, we restore to the classical approach to find the reflectivity of the lattice.

\section{Reflectivity}
\label{sec:methods}
From the coupled dipole equation, Eq. \ref{eq:coupleddipoles}, one can find the dipole moments under a given incident field, 
\begin{equation} \label{eq:dipolesinvM}
    \mathbf{p}(\mathbf{k}_{\|}) = \hat{\mathbf{M}}^{-1}(\omega,\mathbf{k}_{\|}) \mathbf{E}_{\mathrm{inc}} (\mathbf{k}_{\|})\,.
\end{equation}
Here, $\mathbf{p}= \left[\mathbf{p}^0,\dots \mathbf{p}^M\right]$ and similarly for $\mathbf{E}_\mathrm{inc}$.
Thus, we can identify the inverse of the coupled dipole matrix as a \textit{generalized} effective polarizability, 
\begin{eqnarray} \label{eq:generalizedeffalpha}
    \left[ \hat{{\boldsymbol{\beta}}}^{-1}_\text{eff}(\omega,\mathbf{k}_{\|})  \right]^{\mu\nu} =   \frac{\mathbbm{1}}{\alpha^\mu(\omega)} \delta_{\mu,\nu}  -  \frac{k^2}{\varepsilon_0}\hat{\mathbf{S}}^{\mu\nu}(\mathbf{k}_{||}) .
\end{eqnarray}
This \textit{generalized} effective polarisability of the array is a $3M\times3M$ tensor, as corresponds to the 3 spatial coordinates and the $M$ emitters in the unit cell. The effective polarisability tensor is obtained by summing over the sublattice indeces. It is a 3$\times$3 tensor and represents the total response of the array in the cartesian basis. At normal incidence, 
\begin{align}
   \left[ \hat{\boldsymbol{\alpha}}_\text{eff}(\omega,\mathbf{k}_{\|})  \right]= \sum_\mu^M\sum_\nu^M \hat{\boldsymbol{\beta}}^{\mu\nu}_\text{eff}
\end{align}
since the incident field does not introduce extra phases between the different sublattices. 

On the other hand, when considering a reflectivity problem with a plane wave incident on the array, the electric field can be written in the basis of $s$ and $p$ polarisations, with unit vectors, $\mathbf{e}_{p,s}\perp \mathbf{e}_\mathbf{\mathbf{k}_{\|}} $, given by
\begin{eqnarray}
     \mathbf{e}_p^\pm &=& \pm \frac{k_z}{k k_{\|}}   (k_x,k_y, \mp k_{\|}^2/k_z), \\
     \mathbf{e}_s^\pm &=& \frac{1}{k k_{\|}}   (k_y,-k_x,0). 
\end{eqnarray}
Then the incident and reflected fields can be written through their projection over the two polarisations, $ E^\sigma = \mathbf{e}_\sigma\cdot \mathbf{E}(\mathbf{k}_{\|})$,
\begin{eqnarray}
    E^\sigma_\text{inc} = \sum_{\sigma'=p,s} \delta_{\sigma\sigma'}  E^{\sigma'}_{0,\mathbf{k}_{\|}} e^{i\mathbf{k}_{\|}\cdot \mathbf{r}_{\|}} e^{ik_zz}\,, \\
    E^\sigma_\text{ref} = \sum_{\sigma'=p,s} r_{\sigma\sigma'}(\mathbf{k}_{\|})  E^{\sigma'}_{0,\mathbf{k}_{\|}} e^{i\mathbf{k}_{\|}\cdot \mathbf{r}_{\|}} e^{-ik_zz}\,,
\end{eqnarray}
with $r_{\sigma\sigma'}$ being the components of the reflectivity matrix in the polarisation basis, $r_{\sigma\sigma'}=\mathbf{e}_\sigma\cdot\hat{\mathbf{r}}\cdot \mathbf{e}_{\sigma'}$. 

The reflected field is given by the total field radiated by the dipoles in the array after they are excited by the incident wave. From the average transverse effective current generated by the dipoles, $\mathbf{J}_\|= -i\omega \mathbf{p} / A$, with $A$ the unit cell area, we have, 
\begin{equation}    
    \mathbf{E}_{\text{ref},\|}(\mathbf{k}_{\|}) = -\frac{1}{2} Z_0 \mathbf{J}(\mathbf{k}_{\|}) = \frac{i\omega}{2A}Z_0  \mathbf{p}(\mathbf{k}_{\|}) =  \frac{i k }{2A\varepsilon_0}  \mathbf{p}(\mathbf{k}_{\|}) \,,
\end{equation}
where $\mathbf{p}$ stands for the total dipole moment vector in a unit cell which is given by the effective polarisability through Eqs. \ref{eq:dipolesinvM} and \ref{eq:generalizedeffalpha}.

Taking into account the field polarisations, the reflection coefficients are given by, 
\begin{equation}
    r_{\sigma\sigma'}(\mathbf{k}_{\|}) =  \frac{|\mathbf{E}_{\mathrm{ref}}^\sigma(\mathbf{k}_{\|})|}{|\mathbf{E}_{\mathrm{inc}}^{\sigma'}|} = \frac{i k }{2A\varepsilon_0} |\mathbf{e}_\sigma \cdot \hat{\boldsymbol{\alpha}}_\text{eff}\cdot\mathbf{e}_{\sigma'}|
\end{equation}
with incident field $\mathbf{E}_{\mathrm{inc}}^{\sigma'} = \mathbf{E}_0 \mathbf{e}_{\sigma'} $. With this, the array reflectivity is, 
\begin{equation}
    R_{\sigma\sigma'}(\mathbf{k}_{\|}) = |r_{\sigma\sigma'}(\mathbf{k}_{\|})|^2 =  \frac{|\mathbf{e}_\sigma \cdot \hat{\boldsymbol{\alpha}}_\text{eff}\cdot\mathbf{e}_{\sigma'}|^2  }{(2A\varepsilon_0/k)^2} = \frac{|\mathbf{e}_\sigma \cdot \hat{\boldsymbol{\alpha}}_\text{eff}\cdot\mathbf{e}_{\sigma'}|^2 }{(A\lambda\varepsilon_0/\pi)^2}
\end{equation}
For lattices with more than one particle per unit cell, we recall that $\hat{\boldsymbol{\alpha}}_\text{eff}$ represents the total effective polarisability of the unit cell, such that the sublattice indeces are summed over, while the $\mathbf{e}$ vectors act on the spatial indeces of the tensor.

\subsection{Bipartite lattices}

For lattices with 2 particles per cell, the coupled dipole equation reads as,
\begin{equation}  
    \left [\begin{array}{cc}
        \mathbf{p}_1 \\  \mathbf{p}_2
  \end{array}
 \right] 
 = 
    \left [\begin{array}{cc}
  \alpha_1(\omega)^{-1} \mathbbm{1} - (k^2/\varepsilon_0)\hat{\mathbf{S}}_{11}(\omega,\mathbf{k}_{||}) &  - (k^2/\varepsilon_0)\hat{\mathbf{S}}_{12}(\omega,\mathbf{k}_{||})\\   
        - (k^2/\varepsilon_0)\hat{\mathbf{S}}_{21}(\omega,\mathbf{k}_{||}) & \alpha_2(\omega)^{-1} \mathbbm{1} - (k^2/\varepsilon_0)\hat{\mathbf{S}}_{22}(\omega,\mathbf{k}_{||})
    \end{array}
 \right]^{-1} \left [\begin{array}{cc}
        \mathbf{E}_1 \\  \mathbf{E}_2
    \end{array}
 \right].
\end{equation}{} 

The \textit{generalized} effective polarisability in the above matrix equation can be recast as,
\begin{equation}  
    \hat{{\boldsymbol{\beta}}}_\textrm{eff}
 = 3\pi \varepsilon_0 c^3
    \left [\begin{array}{cc}
  \frac{\omega_1^3}{\Gamma_1 } \left((-\omega+\omega_1-i\Gamma_1/2)
   \mathbbm{1} - \frac{\omega^2}{\omega_1^2}\frac{ 3\pi\Gamma_1 c  }{\omega_1 } \hat{\mathbf{S}}_{11}(\omega,\mathbf{k}_{||})\right) &  
   - \frac{\omega_1^3}{\Gamma_1 } \frac{\omega^2}{\omega_1^2}\frac{ 3\pi\Gamma_1 c  }{\omega_1 } \hat{\mathbf{S}}_{12}(\omega,\mathbf{k}_{||})\\   
        - \frac{\omega_2^3}{\Gamma_2  }\frac{\omega^2}{\omega_2^2}\frac{ 3\pi\Gamma_2 c  }{\omega_2 } \hat{\mathbf{S}}_{21}(\omega,\mathbf{k}_{||}) & 
        \frac{\omega_2^3}{\Gamma_2  } \left((-\omega+\omega_2-i\Gamma_2/2)
   \mathbbm{1} - \frac{\omega^2}{\omega_2^2}\frac{ 3\pi\Gamma_2 c  }{\omega_2 } \hat{\mathbf{S}}_{22}(\omega,\mathbf{k}_{||})\right) 
    \end{array}
 \right]^{-1} 
\end{equation}{} 
Given that $|\omega_1-\omega_2|\ll\omega_1$ and linearising, $\omega=\omega_1$, we have, 
\begin{equation}   \label{eq:geneffpolbipartite}
    \hat{{\boldsymbol{\beta}}}_\textrm{eff}
 = \frac{3\pi\Gamma_1 \varepsilon_0 c^3}{\omega_1^3}
    \left [\begin{array}{cc}
  (-\omega+\omega_1-i\Gamma_1/2)
   \mathbbm{1} - \hat{\Tilde{\mathbf{S}}}_{11}(\omega_1,\mathbf{k}_{||})&  
   - \hat{\Tilde{\mathbf{S}}}_{12}(\omega_1,\mathbf{k}_{||})\\   
        - \hat{\Tilde{\mathbf{S}}}_{21}(\omega_1,\mathbf{k}_{||}) & 
        (-\omega+\omega_2-i\Gamma_1/2)
   \mathbbm{1} - \hat{\Tilde{\mathbf{S}}}_{22}(\omega_1,\mathbf{k}_{||})
    \end{array}
 \right]^{-1} 
\end{equation}{} 
where for convenience we have defined, 
\begin{align}
    \hat{\Tilde{\mathbf{S}}} = \frac{ 3\pi\Gamma_1 c  }{\omega_1 } \hat{\mathbf{S}}(\omega_1,\mathbf{k}_{||}) = - \hat{\mathbf{\Delta}} + i \frac{\hat{\mathbf{\Gamma}}}{2}
\end{align}
and we identify $3\pi\Gamma_1 \varepsilon_0 c^3/\omega_1^3 = \alpha_0 \Gamma_1/2 $.

The total dipole moment needed to calculate the array reflectivity is 
\begin{eqnarray}
     p_x^\text{tot} = \sum_{\mu=1}^2\sum_{\nu=1}^2 [\alpha_\text{eff}]_{xx}^{\mu\nu} [E_\text{inc}]_{x}^{\mu} =  
        p_x^1+ p_x^2\,,
\end{eqnarray}
with 
\begin{eqnarray}
    p_x^1 &=&  [\beta_\text{eff}]_{xx}^{11} [E_\text{inc}]_{x}^{1} + [\beta_\text{eff}]_{xx}^{12} [E_\text{inc}]_{x}^{2}\,, \\
    p_x^2 &=&  [\beta_\text{eff}]_{xx}^{21} [E_\text{inc}]_{x}^{1} + [\beta_\text{eff}]_{xx}^{22} [E_\text{inc}]_{x}^{2},
\end{eqnarray}
where at normal incidence we have $ [E_\text{inc}]_{x}^{1} = [E_\text{inc}]_{x}^{2} = E_0$, and the \textit{generalized} effective polarisability is given by Eq. \ref{eq:geneffpolbipartite}

We now write explicitly the effective polarisability tensor of the bipartite square lattice. We first note that the equivalence of the two sublattices implies $\hat{\mathbf{S}}_{11} = \hat{\mathbf{S}}_{22} $. Symmetry of the square lattice and normal incidence imply $S^{11}_{xx} = S^{11}_{yy}$, $S^{11}_{xy} = S^{11}_{yx}=0$ and $\hat{\mathbf{S}}_{12}(\mathbf{k}_\|=0) = \hat{\mathbf{S}}_{21}(\mathbf{k}_\|=0) $. In general, the other lattice sum elements are different from zero. We consider the $x,y$ components, which are decoupled from the $z$ ones, 
\begin{equation}  
    \hat{{\boldsymbol{\beta}}}_\textrm{eff}
    = \alpha_0\frac{\Gamma_1}{2}
    \begin{bmatrix}
        -\omega+\omega_1-i\Gamma_1/2 - \tilde{S}^{11}_{xx} & 0 & - \tilde{S}^{12}_{xx} & - \tilde{S}^{12}_{xy} \\
      0 & -\omega+\omega_1-i\Gamma_1/2- \tilde{S}^{11}_{xx} & - \tilde{S}^{12}_{xy} & - \tilde{S}^{12}_{yy} \\ 
        - \tilde{S}^{12}_{xx} & - \tilde{S}^{12}_{xy} & -\omega+\omega_"-i\Gamma_2/2 - \tilde{S}^{11}_{xx} & 0 \\
        - \tilde{S}^{12}_{xy} & - \tilde{S}^{12}_{yy} & 0 & -\omega+\omega_2-i\Gamma_2/2 - \tilde{S}^{11}_{xx}
    \end{bmatrix}^{-1}
\end{equation}{} 
The above can also be rearranged as, 
\begin{equation}  
    \hat{{\boldsymbol{\beta}}}_\textrm{eff}
    = \alpha_0\frac{\Gamma_1}{2}
    \begin{bmatrix}
        -\omega+\omega_1-i\Gamma_1/2 - \tilde{S}^{11}_{xx} & - \tilde{S}^{12}_{xx} & 0 & - \tilde{S}^{12}_{xy} \\ - \tilde{S}^{12}_{xx} & -\omega+\omega_2-i\Gamma_2/2 - \tilde{S}^{11}_{xx} & - \tilde{S}^{12}_{xy}  & 0 \\
      0 & - \tilde{S}^{12}_{xy} & -\omega+\omega_1-i\Gamma_1/2- \tilde{S}^{11}_{xx}  & - \tilde{S}^{12}_{yy} \\ 
        - \tilde{S}^{12}_{xy} & 0 & - \tilde{S}^{12}_{yy}  & -\omega+\omega_2-i\Gamma_2/2 - \tilde{S}^{11}_{xx}
    \end{bmatrix}^{-1}
\end{equation}{} 
representing $\{xx,xy,yx,yy\}$ blocks instead of $\{11,12,21,22\}$ ones.

In general, the above expression tells us that the interaction between the two sublattices will couple the two polarisations. However, if the second sublattice is placed at a high symmetry point, we can simplify more. For instance, for a square lattice, if the second sublattice is placed on one of the mirror lines of the lattice (horizontal, vertical or diagonal), we have $S^{12}_{xy} = 0$ and the problem block diagonalises, each block corresponding to one polarisation. Furthermore, along the diagonal mirror line, we also have $S^{12}_{xx} = S^{12}_{yy}$, and 
\begin{equation}  \label{eq:betaeffdiag}
    \hat{{\boldsymbol{\beta}}}_\textrm{eff}
    = \alpha_0\frac{\Gamma_1}{2}
    \begin{bmatrix}
        -\omega+\omega_1-i\Gamma_1/2 - \tilde{S}^{11}_{xx} & - \tilde{S}^{12}_{xx} & 0 & 0\\ - \tilde{S}^{12}_{xx} & -\omega+\omega_2-i\Gamma_2/2 - \tilde{S}^{11}_{xx} & 0 & 0 \\
      0 & 0 & -\omega+\omega_1-i\Gamma_1/2- \tilde{S}^{11}_{xx}  & - \tilde{S}^{12}_{xx} \\ 
        0 & 0 & - \tilde{S}^{12}_{xx}  & -\omega+\omega_2-i\Gamma_2/2 - \tilde{S}^{11}_{xx}
    \end{bmatrix}^{-1}.
\end{equation}{} 
Hence in this case the two polarisations are completely equivalent at normal incidence, as expected due to the symmetry of the lattice. For one polarisation,
\begin{equation}  
    \begin{bmatrix}
        p^1_x \\ p^2_x  
  \end{bmatrix}
    = \alpha_0\frac{\Gamma_1}{2}
    \begin{bmatrix}
        -\omega+\omega_1-i\Gamma_1/2- \tilde{S}^{11}_{xx} & - \tilde{S}^{12}_{xx} \\   
      - \tilde{S}^{12}_{xx} & -\omega+\omega_2-i\Gamma_2/2 - \tilde{S}^{11}_{xx}  
    \end{bmatrix}^{-1}
    \begin{bmatrix}
        E^1_x \\   E^2_x 
    \end{bmatrix},
\end{equation}{} 
and the \textit{generalized} effective polarizability tensor is, 
\begin{eqnarray} 
    \hat{\boldsymbol{\beta}}^{xx}_\text{eff} = \alpha_0\frac{\Gamma_1}{2} \begin{bmatrix}
        -\omega+\omega_1-i\Gamma_1/2- \tilde{S}^{11}_{xx} & - \tilde{S}^{12}_{xx} \\   
      - \tilde{S}^{12}_{xx} & -\omega+\omega_2-i\Gamma_2/2 - \tilde{S}^{11}_{xx}   \\
    \end{bmatrix}^{-1}
\end{eqnarray}

At normal incidence, the dipole moments in each sublattice are,
\begin{eqnarray}
    p_x^1 &=& \alpha_0\frac{\Gamma_1}{2}  \frac{-\omega+\omega_2-i\Gamma_2/2+ (\tilde{S}^{12}_{xx}-\tilde{S}^{11}_{xx})}{\left(-\omega+\omega_1-i\Gamma_1/2-\tilde{S}^{11}_{xx}\right) \left(-\omega+\omega_2-i\Gamma_2/2-\tilde{S}^{11}_{xx}\right)-\left(\tilde{S}^{12}_{xx}\right)^2} E_0, \\
    p_x^2 &=& \alpha_0\frac{\Gamma_1}{2} \frac{-\omega+\omega_1-i\Gamma_1/2+ (\tilde{S}^{12}_{xx}-\tilde{S}^{11}_{xx})}{\left(-\omega+\omega_1-i\Gamma_1/2-\tilde{S}^{11}_{xx}\right) \left(-\omega+\omega_2-i\Gamma_2/2-\tilde{S}^{11}_{xx}\right)-\left(\tilde{S}^{12}_{xx}\right)^2}E_0,
\end{eqnarray}
and the total dipole moment, 
\begin{eqnarray}
    p_x^{tot} &=& \alpha_0\frac{\Gamma_1}{2} \frac{-2\omega+\omega_1+\omega_2-i\Gamma_1/2-i\Gamma_2/2+ 2(\tilde{S}^{12}_{xx}-\tilde{S}^{11}_{xx})}{ \left(-\omega+\omega_1-i\Gamma_1/2-\tilde{S}^{11}_{xx}\right) \left(-\omega+\omega_2-i\Gamma_2/2-\tilde{S}^{11}_{xx}\right)-(\tilde{S}^{12}_{xx})^2} E_0,
\end{eqnarray}
where the lattice sums are evaluated at $\mathbf{k}_\| = 0 $.

Thus, the effective polarisability of the array is, 
\begin{eqnarray}
    \alpha^{xx}_\textrm{eff}  &=& \alpha_0\frac{\Gamma_1}{2} \frac{- (\delta_1 + i\Gamma_1/2+\delta_2 + i\Gamma_2/2)+ 2(\tilde{S}^{12}_{xx}-\tilde{S}^{11}_{xx})}{ \left(\delta_1+i\Gamma_1/2+\tilde{S}^{11}_{xx}\right) \left(\delta_2+i\Gamma_2/2+\tilde{S}^{11}_{xx}\right)-(\tilde{S}^{12}_{xx})^2}, 
\end{eqnarray}
and the co-polarised reflectivity, $R_{pp}=R_{ss} $, 
\begin{equation}
        R_{pp}(\mathbf{k}_{\|}=0) =   \left( \frac{\alpha_0\Gamma_1 }{2A\lambda\varepsilon_0/\pi}\right)^2 \left|\frac{- (\delta_1 + i\Gamma_1/2+\delta_2 + i\Gamma_2/2+ 2(\tilde{S}^{12}_{xx}-\tilde{S}^{11}_{xx})}{ \left(\delta_1+i\Gamma_1/2+\tilde{S}^{11}_{xx}\right) \left(\delta_2+i\Gamma_2/2+\tilde{S}^{11}_{xx}\right)-(\tilde{S}^{12}_{xx})^2} \right|^2
\end{equation}

In order to look for the resonances of the effective polarisability (and thus of the reflectivity spectrum) it is useful to express it using new variables, 
\begin{align}
    \delta_+ &= -(\delta_1+\delta_2)/2 = - \omega + (\omega_1 + \omega_2) / 2  \\
    \delta_- &= -(\delta_1-\delta_2)/2 = (\omega_1 - \omega_2) / 2 ,
\end{align}
which yields, 
\begin{align}
    \alpha^{xx}_\textrm{eff} &= \alpha_0\Gamma_1  \frac{ \delta_+ - i\Gamma/2 + \tilde{S}^{12}_{xx}-\tilde{S}^{11}_{xx}}{ \left(\delta_+ - \delta_- -\tilde{S}^{11}_{xx} - i\Gamma_1/2 \right) \left(\delta_+ + \delta_- -\tilde{S}^{11}_{xx} - i\Gamma_1/2 \right)-(\tilde{S}^{12}_{xx})^2}  \\ &=  \alpha_0\Gamma_1 \frac{\delta_+ - i\Gamma/2 + \tilde{S}^{12}_{xx}-\tilde{S}^{11}_{xx} }{ (\delta_+ - \Omega_+ ) (\delta_+ - \Omega_- )} ,
\end{align}
where we have assumed $\Gamma_1\approx\Gamma_2$, given that the decay rates are much smaller than the frequencies, and introduced the quantities,
\begin{align}
    \Omega_\pm =  \tilde{S}^{11}_{xx} \pm \textrm{sign}\left(-\textrm{Re}(\tilde{S}^{12}_{xx})\right) \sqrt{\left(\tilde{S}^{12}_{xx}\right)^2 + (\delta_-)^2 } +i\Gamma/2.
\end{align}
Then, we can write, 
\begin{align}
    \alpha^{xx}_\textrm{eff} &=  \alpha_0\Gamma_1 \left[ \frac{ \Sigma_+ }{ \delta_+ - \Omega_+ }  + \frac{ \Sigma_- }{ \delta_+ - \Omega_- } \right],
\end{align}
with, 
\begin{align}
    \Sigma_\pm=   \frac{1}{2}  \pm  \textrm{sign}\left(-\textrm{Re}(\tilde{S}^{12}_{xx})\right) \frac{\tilde{S}^{12}_{xx} }{ 2 \sqrt{\left(\tilde{S}^{12}_{xx}\right)^2 + (\delta_-)^2 }}  \\
\end{align}
This expression was derived for a bipartite array of plasmonic nanoparticles in Ref. \cite{acsnanosupersub}. 
\begin{figure}[b!]
  \centering
  \includegraphics[width=0.6\textwidth]{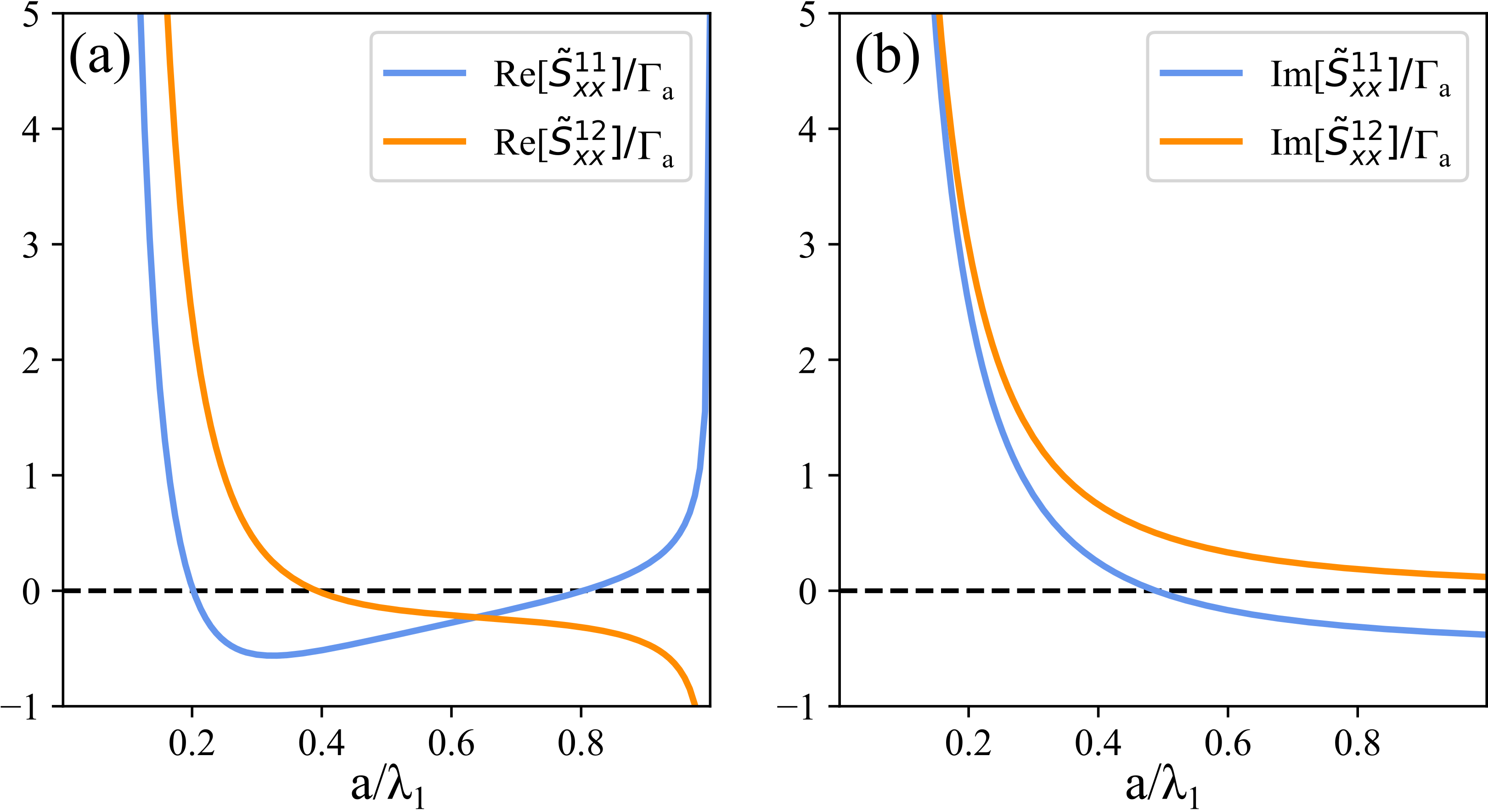}
  \caption{Relevant components of the real and imaginary parts of the lattice sums $\hat{\overline{\textbf{S}}}$, that is, the cooperative shift and width, as function of lattice periodicity for the bipartite case considered in Fig. 2 of the main text and at normal incidence ($\mathbf{k}_{||} = 0$). Panel (a) shows the real parts of $\tilde{S}^{11}_{xx}/\Gamma_a$ depicted in blue, and $\tilde{S}^{12}_{xx}/\Gamma_a$ in orange, while the corresponding imaginary parts are shown in panel (b). }
  \label{figsuplattice}
\end{figure}
In the emitter array, the only frequency dependence in these expressions is through $\delta_+ = -\omega +(\omega_1+\omega_2)/2$, 
\begin{align}
   \alpha^{xx}_\textrm{eff}(\omega,\mathbf{k}_{||}) &=  \alpha_0\Gamma_1 \left[ \frac{ \Sigma_+(\mathbf{k}_{||}) }{ \delta_+(\omega) - \Omega_+(\mathbf{k}_{||}) }  + \frac{ \Sigma_ -(\mathbf{k}_{||}) }{ \delta_+(\omega) - \Omega_-(\mathbf{k}_{||}) } \right],
    \end{align}
while the other quantities depend only on the Bloch wavevector through the linearised lattice sums, 
\begin{align}
     \Omega_\pm (\mathbf{k}_{||}) &=  -\Delta_{xx}^{11}(\mathbf{k}_{||}) + i \frac{\Gamma^{11}_{xx}(\mathbf{k}_{||})}{2} \pm \textrm{sign} \left( -\Delta_{xx}^{12} \right) \sqrt{\left(-\Delta_{xx}^{12}(\mathbf{k}_{||}) + i \frac{\Gamma^{12}_{xx}(\mathbf{k}_{||})}{2}\right)^2 + \left(\frac{\omega_1-\omega_2}{2}\right)^2 } +i\frac{\Gamma_1}{2}, \\
      \Sigma_\pm (\mathbf{k}_{||}) &=   \frac{1}{2} \left( 1 \pm  \textrm{sign} \left( -\Delta_{xx}^{12} \right) \frac{-\Delta_{xx}^{12}(\mathbf{k}_{||}) + i \Gamma^{12}_{xx}(\mathbf{k}_{||})/2 }{  \sqrt{\left(-\Delta_{xx}^{12}(\mathbf{k}_{||}) + i \Gamma^{12}_{xx}(\mathbf{k}_{||})/2\right)^2 + (\omega_1-\omega_2)^2/4  }} \right)
\end{align}
Equivalently, 
\begin{align}
     \Omega_\pm (\mathbf{k}_{||}) &=  -\Delta_{xx}^{11}(\mathbf{k}_{||}) \pm \Lambda(\mathbf{k}_{||}) + i \frac{\Gamma^{11}_{xx}(\mathbf{k}_{||})}{2} +i\frac{\Gamma_1}{2}, \\
      \Sigma_\pm (\mathbf{k}_{||}) &=   \frac{1}{2} \left( 1 \pm   \frac{-\Delta_{xx}^{12}(\mathbf{k}_{||}) + i \Gamma^{12}_{xx}(\mathbf{k}_{||})/2 }{  \Lambda(\mathbf{k}_{||})} \right)
\end{align}

with 
\begin{align}
    \Lambda(\mathbf{k}_{||})=\textrm{sign} \left( -\Delta_{xx}^{12} \right) \sqrt{\left(-\Delta_{xx}^{12}(\mathbf{k}_{||}) + i \frac{\Gamma^{12}_{xx}(\mathbf{k}_{||})}{2}\right)^2 + \left(\frac{\omega_1-\omega_2}{2}\right)^2 } 
\end{align}

The real and imaginary parts of the lattice sums, which correspond to the cooperative shift and decay, respectively, are depicted as function of the lattice periodicity in Fig.~\ref{figsuplattice}. We see how $\text{Re}[S_{xx}^{12}] $ changes from positive to negative as the period increases, which explains the frequency ordering of the subradiant and superradiant modes as function of the period shown in Fig. 3 of the main text. EIT arises precisely when $\text{Re}[S_{xx}^{12}]=0 $ since the modes become degenerate, as can be seen from Eq. \ref{eq:betaeffdiag}.

\subsubsection{Zero detuning} 
When the two sublattices are not detuned, $\omega_2=\omega_1$, we have $\delta_-=0$, $\delta_+ = - \delta_1$, $\Sigma_- = 0 $, $\Sigma_+ = 1 $. Therefore, 
\begin{align}
    \alpha^{xx}_\textrm{eff} &=  \alpha_0\frac{\Gamma_1}{2}  \frac{ \Sigma_+ }{ \delta_+ - \Omega_+ }  = -\alpha_0 \frac{ \Gamma_1/2 }{ \delta_1 - \Delta_{xx}^{11}(\mathbf{k}_{||})  -\Delta_{xx}^{12}(\mathbf{k}_{||})  + i \left(\Gamma_1 + \Gamma^{11}_{xx}(\mathbf{k}_{||}) + \Gamma^{12}_{xx}(\mathbf{k}_{||})\right)/2  } ,
\end{align}
which results in a Lorentzian response and we can identify scalar functions as cooperative shift, $\Delta(\mathbf{k}_{||})=\Delta_{xx}^{11}(\mathbf{k}_{||})  -\Delta_{xx}^{12}(\mathbf{k}_{||})$, and decay, $ \Gamma(\mathbf{k}_{||}) = \Gamma^{11}_{xx}(\mathbf{k}_{||}) -\Gamma^{12}_{xx}(\mathbf{k}_{||})$. 

\subsubsection{Going back to the Bravais lattice}
For the case of one particle per cell, all the inter-lattice lattice sums vanish and we recover the usual result, 
\begin{align}
    \alpha^{xx}_\textrm{eff} &=  - \alpha_0 \frac{ \Gamma_1/2 }{ \omega - \omega_1 - \Delta_{xx}^{11}(\mathbf{k}_{||})   + i \left(\Gamma_1 + \Gamma^{11}_{xx}(\mathbf{k}_{||})\right)/2  } ,
\end{align}
where $\Delta_{xx}^{11}$ and $\Gamma^{11}_{xx} $ correspond to the cooperative shift and cooperative decay, respectively. 

\begin{figure}[ht]
  \centering
  \includegraphics[width=0.9\textwidth]{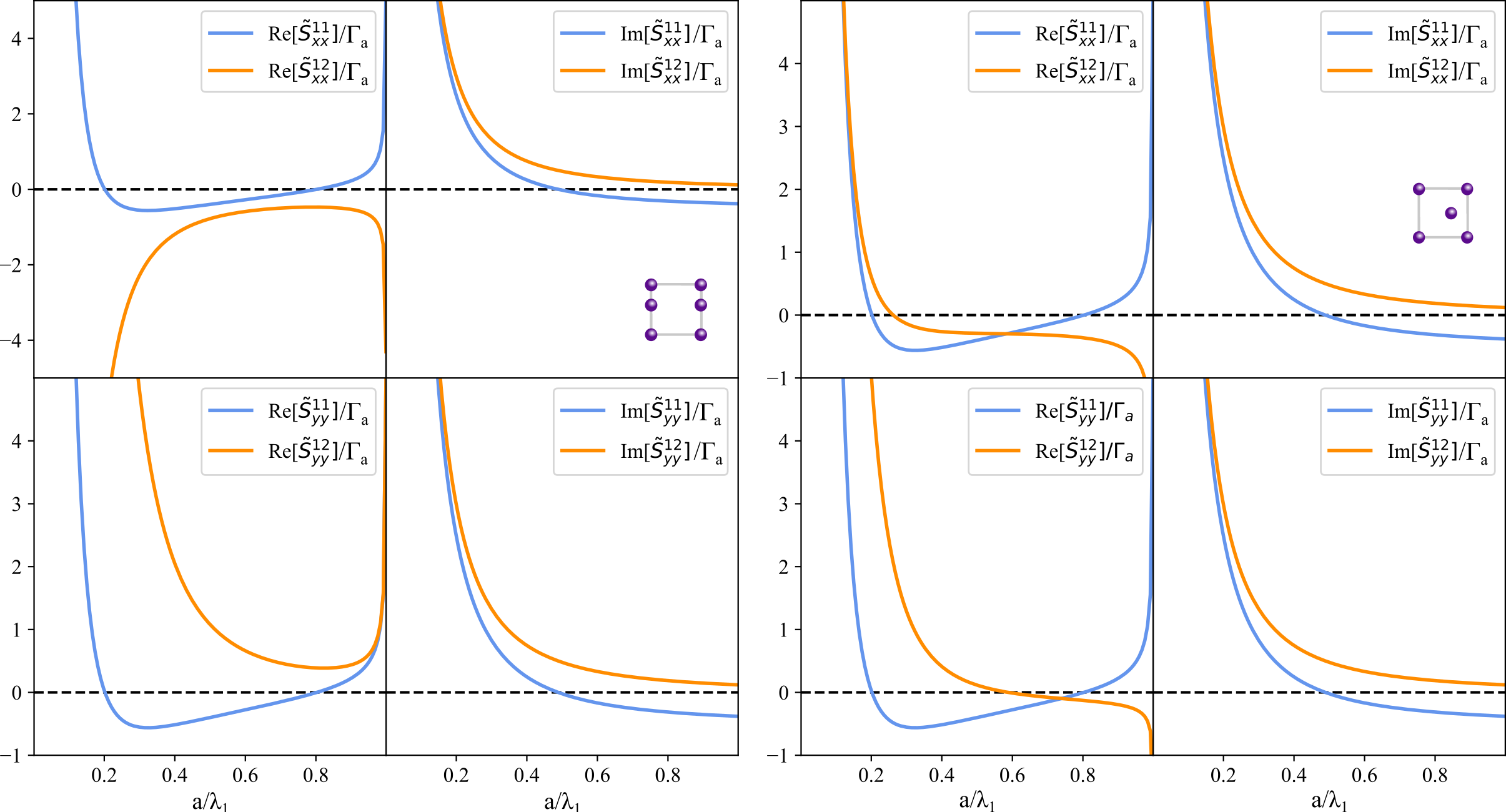}
  \caption{Real and imaginary part of the lattice sums at normal incidence as function of the lattice period $a/\lambda_1$: $xx$-components in the top part, and $yy$-components in the lower part. Intra-sublattice terms are depicted in blue while inter-sublattice ones appear in orange. The left part corresponds to a lattice with basis vector $T=(0,0.6)d$, while the right part to a lattice with basis vector $T=(0.7,0.5)d$. }
  \label{SMlat}
\end{figure}

\subsection{Designing a polariser: Bipartite lattice with different response to $s-$ and $p-$ polarisations}

    Here we consider a different geometry that breaks the symmetry between $s-$ and $p-$ polarisations. 
    For that purpose we consider a square array where the second sublattice is placed on on a horizontal or vertical mirror line. This breaks the diagonal mirror symmetry of the lattice studied in the previous section, resulting in different response to different polarisations, but still avoids cross-polarised terms.  
    Again, we write explicitly the effective polarisability tensor. At normal incidence the symmetry of the lattice implies $S^{11}_{xx} = S^{11}_{yy}$, $S^{11}_{xy} = S^{11}_{yx}=0$ and
$S^{12}_{xx} = S^{21}_{xx} \neq S^{12}_{yy} = S^{21}_{yy}$, $S^{12}_{xy} = S^{12}_{yx}=0$. Hence in this case we still have a block diagonal matrix,
\begin{equation}  
    \hat{{\boldsymbol{\beta}}}_\textrm{eff}
    = \alpha_0\frac{\Gamma_1}{2}
    \begin{bmatrix}
        -\omega+\omega_1-i\Gamma_1/2 - \tilde{S}^{11}_{xx} & - \tilde{S}^{12}_{xx} & 0 & 0 \\ - \tilde{S}^{12}_{xx} & -\omega+\omega_2-i\Gamma_2/2 - \tilde{S}^{11}_{xx} & 0  & 0 \\
      0 & 0 & -\omega+\omega_1-i\Gamma_1/2- \tilde{S}^{11}_{yy}  & - \tilde{S}^{12}_{yy} \\ 
        0 & 0 & - \tilde{S}^{12}_{yy}  & -\omega+\omega_2-i\Gamma_2/2 - \tilde{S}^{11}_{yy}
    \end{bmatrix}^{-1}
\end{equation}{} 
representing $\{xx,yy\}$ blocks that are accessed by light of different polarization. We can now write for the two polarisations,
\begin{eqnarray} 
    \hat{\boldsymbol{\beta}}^{xx}_\text{eff} = \alpha_0\frac{\Gamma_1}{2} \begin{bmatrix}
        -\omega+\omega_1-i\Gamma_1/2- \tilde{S}^{11}_{xx} & - \tilde{S}^{12}_{xx} \\   
      - \tilde{S}^{12}_{xx} & -\omega+\omega_2-i\Gamma_2/2 - \tilde{S}^{11}_{xx}   \\
    \end{bmatrix}^{-1}
\end{eqnarray}
and
\begin{eqnarray} 
    \hat{\boldsymbol{\beta}}^{yy}_\text{eff} = \alpha_0\frac{\Gamma_1}{2} \begin{bmatrix}
        -\omega+\omega_1-i\Gamma_1/2- \tilde{S}^{11}_{yy} & - \tilde{S}^{12}_{yy} \\   
      - \tilde{S}^{12}_{yy} & -\omega+\omega_2-i\Gamma_2/2 - \tilde{S}^{11}_{yy}   \\
    \end{bmatrix}^{-1}
\end{eqnarray}

Thus, the effective polarisability of the array along the $x-$direction is, 
\begin{eqnarray}
    \alpha^{xx}_\textrm{eff}  &=& \alpha_0\frac{\Gamma_1}{2} \frac{- (\delta_1 + i\Gamma_1/2+\delta_2 + i\Gamma_2/2)+ 2(\tilde{S}^{12}_{xx}-\tilde{S}^{11}_{xx})}{ \left(\delta_1+i\Gamma_1/2+\tilde{S}^{11}_{xx}\right) \left(\delta_2+i\Gamma_2/2+\tilde{S}^{11}_{xx}\right)-(\tilde{S}^{12}_{xx})^2}, 
\end{eqnarray}
while along the $y-$direction, 
\begin{eqnarray}
    \alpha^{yy}_\textrm{eff}  &=& \alpha_0\frac{\Gamma_1}{2} \frac{- (\delta_1 + i\Gamma_1/2+\delta_2 + i\Gamma_2/2)+ 2(\tilde{S}^{12}_{yy}-\tilde{S}^{11}_{yy})}{ \left(\delta_1+i\Gamma_1/2+\tilde{S}^{11}_{yy}\right) \left(\delta_2+i\Gamma_2/2+\tilde{S}^{11}_{yy}\right)-(\tilde{S}^{12}_{yy})^2}.
\end{eqnarray}
Therefore, the polarisation-dependent reflectivity, $R_{pp}\neq R_{ss}$, is given by
\begin{equation}
        R_{ss}(\mathbf{k}_{\|}=0) =   \left( \frac{\alpha_0\Gamma_1 }{2A\lambda\varepsilon_0/\pi}\right)^2 \left|\frac{- (\delta_1 + i\Gamma_1/2+\delta_2 + i\Gamma_2/2+ 2(\tilde{S}^{12}_{xx}-\tilde{S}^{11}_{xx})}{ \left(\delta_1+i\Gamma_1/2+\tilde{S}^{11}_{xx}\right) \left(\delta_2+i\Gamma_2/2+\tilde{S}^{11}_{xx}\right)-(\tilde{S}^{12}_{xx})^2} \right|^2
\end{equation}
\begin{equation}
        R_{pp}(\mathbf{k}_{\|}=0) =   \left( \frac{\alpha_0\Gamma_1 }{2A\lambda\varepsilon_0/\pi}\right)^2 \left|\frac{- (\delta_1 + i\Gamma_1/2+\delta_2 + i\Gamma_2/2+ 2(\tilde{S}^{12}_{yy}-\tilde{S}^{11}_{xx})}{ \left(\delta_1+i\Gamma_1/2+\tilde{S}^{11}_{xx}\right) \left(\delta_2+i\Gamma_2/2+\tilde{S}^{11}_{xx}\right)-(\tilde{S}^{12}_{yy})^2} \right|^2
\end{equation}
Then, in this case, the response of the lattice to light of different polarisation can be very different, as it is strongly influenced by the directional interaction among sublattices. Fig.~\ref{SMlat} shows the real and imaginary part of the lattice sums for the lattices as function of the lattice period, for the lattices used in section~\ref{transSM} and the main text. The blue line denotes the intra-sublattice interactions while the orange one represents inter-sublattices interactions. The top panels show the $xx-$components of the lattice sums, corresponding to $s-$polarized wave excitation, while the bottom panels show the $yy-$components corresponding to $p-$polarized waves.   

\section{Retardation effects}

\begin{figure}[t!]
  \centering
  \includegraphics[width=\textwidth]{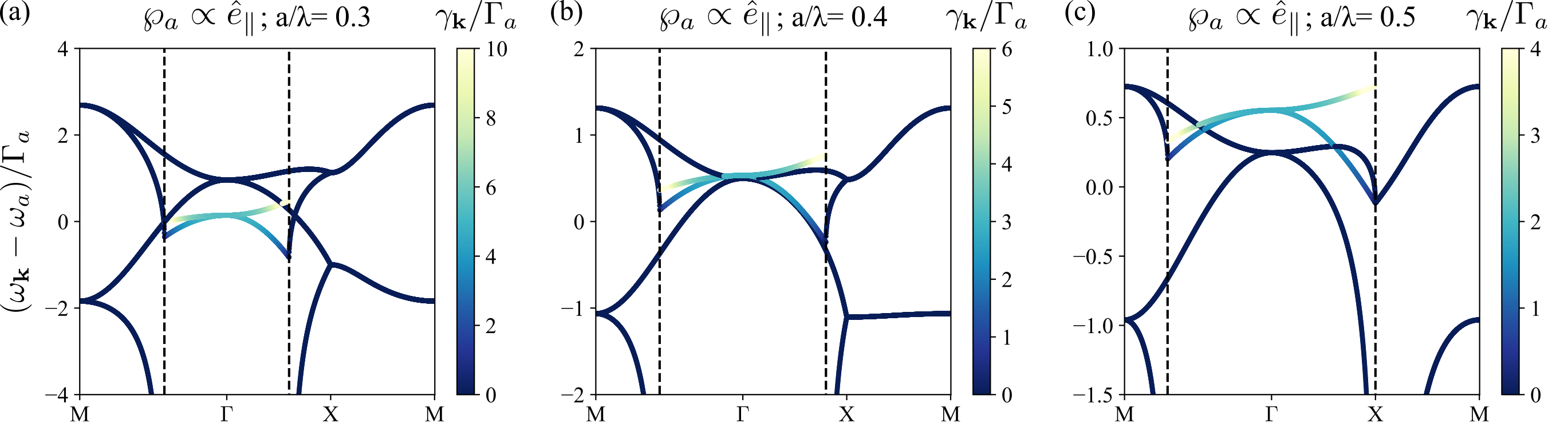}
  \caption{Band structures of bipartite square lattice with $\mathbf{T}=(0.5,0.5)a$ and $\omega_{1,2}=\omega_a$ for increasing lattice periodicity: (a) $a=0.3\lambda_a$, (b) $a=0.4\lambda_a$ , (c) $a=0.5\lambda_a$. }
  \label{figSMbands}
\end{figure}

Here we present complementary band structures to the results shown in Fig. 2 of the main text, where we considered a bipartite lattice with periodicity $a/\lambda_a = 0.2$.  Fig.~\ref{figSMbands} presents the bands for larger values of lattice periodicit. This completes the study of the crossing between the subradiant and superradiant modes discussed in the main text for normal incidence. We observe that for period $a/\lambda_a = 0.3$ the subradiant modes appear at higher frequency than the superradiant ones, as shown in panel (a). For periodicity $a/\lambda_a = 0.39$ a four-fold accidental degeneracy appears at $\mathbf{\Gamma}$ -- bands for $a/\lambda_a = 0.4$ are shown in panel (b). For longer periodicities the subradiant modes slightly decrease in frequency, while the superradiant modes move to higher frequencies, as shown for $a/\lambda_a = 0.5$ in panel (c). We also observe how as the period increases the light line has a stronger effect on the modes. However, being at the center of the zone, the BICs are not affected by retardation effects. 
%
\begin{figure}[t!]
  \centering
  \includegraphics[width=\textwidth]{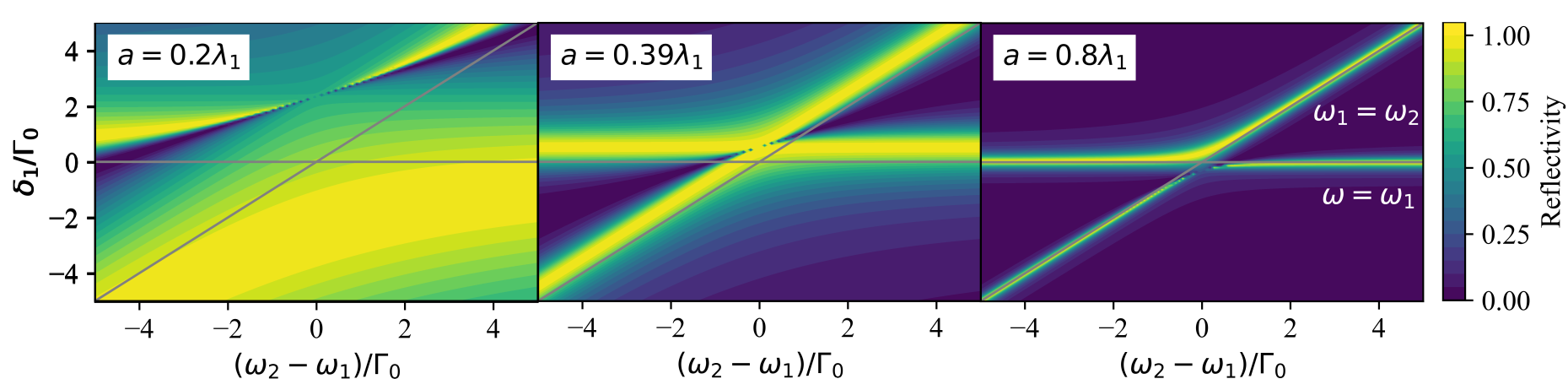}
  \caption{Reflectivity $R$ for bipartite lattices of periods $a/\lambda_{a,1}=0.2\,,0.39\,,0.8$, from top to bottom. $R$ is plotted as a function of frequency, parametrized with detuning of first atom, $\delta_1=\omega-\omega_1$, and the detuning between the two atoms, $\omega_2-\omega_1$. The horizontal gray lines denote zero detuning, $\omega=\omega_1$, while the diagonal ones denote equal atoms, $\omega_2=\omega_1$.}
  \label{figSMAnticrossings}
\end{figure}

Additionally, we also show a study of the anticrossing behaviour of the subradiant and superradiant modes. 
The reflectivity of the lattice for periodicities $a/\lambda_a = 0.2, 0.39, 0.8$, is depicted in Fig.~\ref{figSMAnticrossings} as function of the detuning of the incident light with respect the first emitter and the detuning among the emitters of the two different sublattices. 
From the countour plots we can observe the avoided corssing between the broad, superradiant modes and the narrow, subradiant modes. For $a/\lambda_a = 0.2$ (left) and $a/\lambda_a = 0.8$ (right), the BICs can be observed as vanishing reflectivity for zero detuning between the two sublattices, with asymmetric Fano peaks appearing for finite detuning. On the other hand, when $a/\lambda_a = 0.39$ (center), the subradiant and superradiant modes overlap and we observe an anticrossing between two modes of equal width: this is the case of EIT, identified by the zero reflectivity point at zero detuning.

\section{Electromagnetic induced transparency}
In the main text we discussed the EIT behaviour for the lattice respecting the diagonal mirror symmetry. At a given value of periodicity we have $\text{Re}[S_{xx}^{12}=0]$, and as a consequence the superradiant and the quasi-BIC modes are degenerate, resulting in an EIT window. For the lattice considered in the main text this occurs at $a/\lambda_a = 0.39$ for $\omega_2=\omega_1+\Gamma_a/2$ (Fig. 3b). For this geometry, we have seen that symmetry implies $S_{xx}^{\mu\nu}=S_{yy}^{\mu\nu}$ such that the two polarisations are equivalent. As a consequence, four modes are degenerate at the EIT condition, two BICs corresponding to the $x$ and $y$ degrees of freedom, and two superradiant modes corresponding to the $x$ and $y$. Here we explore how the breaking of the lattice diagonal mirror symmetry  results in two EIT windows at different frequencies, one for each polarisation. 

We start from the configuration presented in the main text (Figs. 2-4), and consider a basis vector that displaces the second sublattice along the horizontal direction, $\textbf{T}=a/2(1,1)+(x,0)a$. This breaks the diagonal mirror symmetry, implying $S_{xx}^{\mu\nu}\neq S_{yy}^{\mu\nu}$, such that the four-fold degeneracy splits into two pairs of modes: the quasi-BIC and the superadiant mode pair corresponding to the $x$ direction split from the pair corresponding to the $y$ direction. Thus, the EIT behaviour is conserved for each polarisation separately. 

\begin{figure}[ht]
  \centering
  \includegraphics[width=0.6\textwidth]{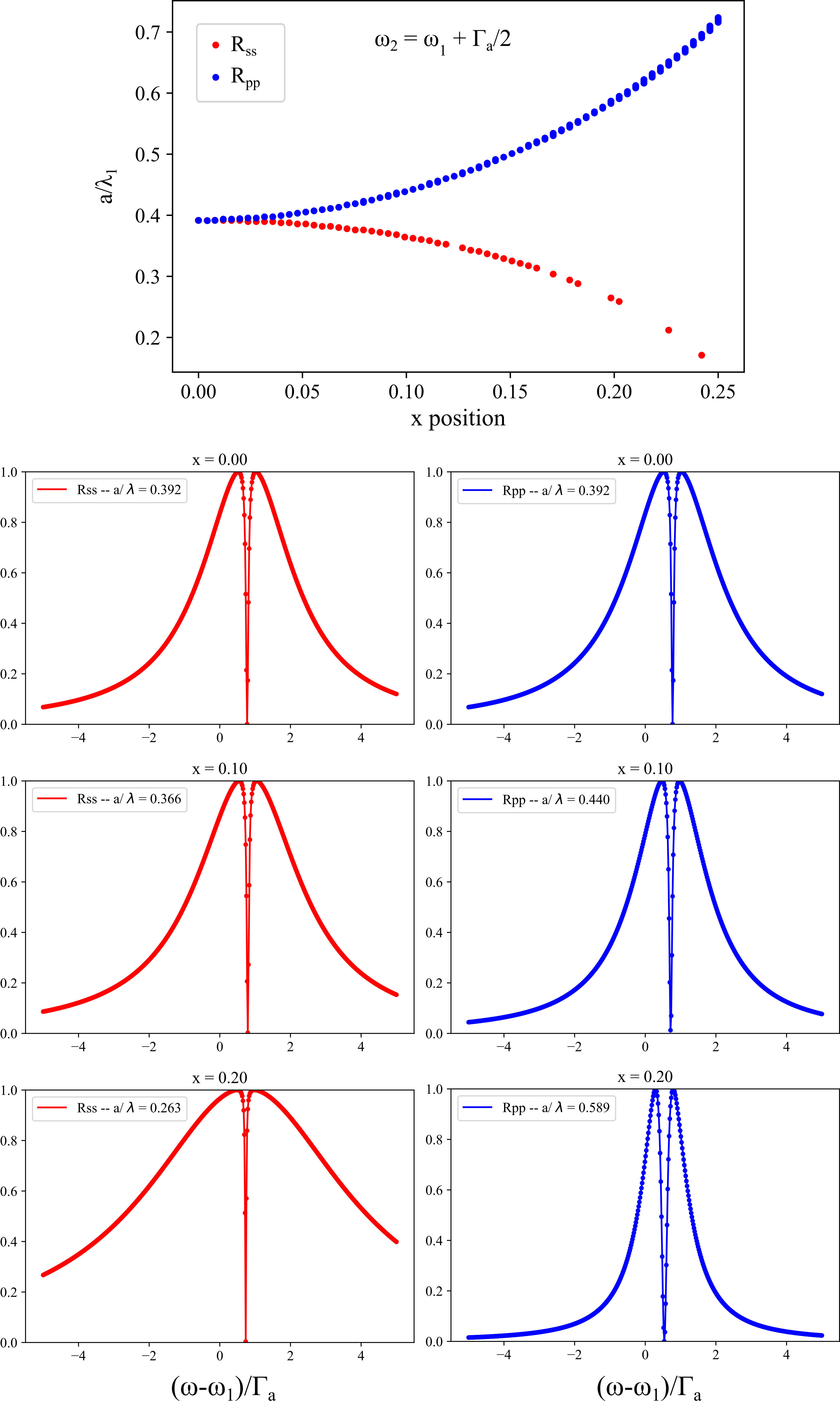}
  \caption{Response of a bipartite lattice with basis vector $\textbf{T}=(1,1)a/2+(x,0)a$ and emitters with detuned frequencies $\omega_2 = \omega_1 + \Gamma_a /2$. The top panel shows the lattice periodicity at which EIT is found for each polarisation as function of $x$, the displacement of the second sublattice. 
  The red and blue dots correspond to the resonance condition for $s-$ and $p-$polarizations, respectively. Reflectivity spectra are shown below are plotted, for different values of the lattice period and size of the basis vector, $x$, for the two polarizations. }
  \label{EIT}
\end{figure}
The EIT window for $s-$ and $p-$polarisations can be tracked by $\text{Re}[S_{xx}^{12}]=0$ and $\text{Re}[S_{yy}^{12}]=0$, respectively, which shift to lower and higher values of the periodicity as the second sublattice is displaced more and more along $x$. 
Figure~\ref{EIT} (top) displays the EIT periodicity for the two polarisations as the second sublattice is displaced. 
Increasing the size of the basis vector along $x$, we observe that for incident $s-$polarized waves, the EIT response appears at lower values of the period, while for $p-$polarization it appears at larger periods. 
The panels below display reflectivity spectra for $s-$ and $p-$polarizations (left and right columns, respectively), for the values of lattice period that display EIT as the basis vector size increases (from top to bottom). As can be seen, an EIT window can be found for both polarizations.

\section{Transmission}\label{transSM}
\begin{figure}[ht]
  \centering
  \includegraphics[width=\textwidth]{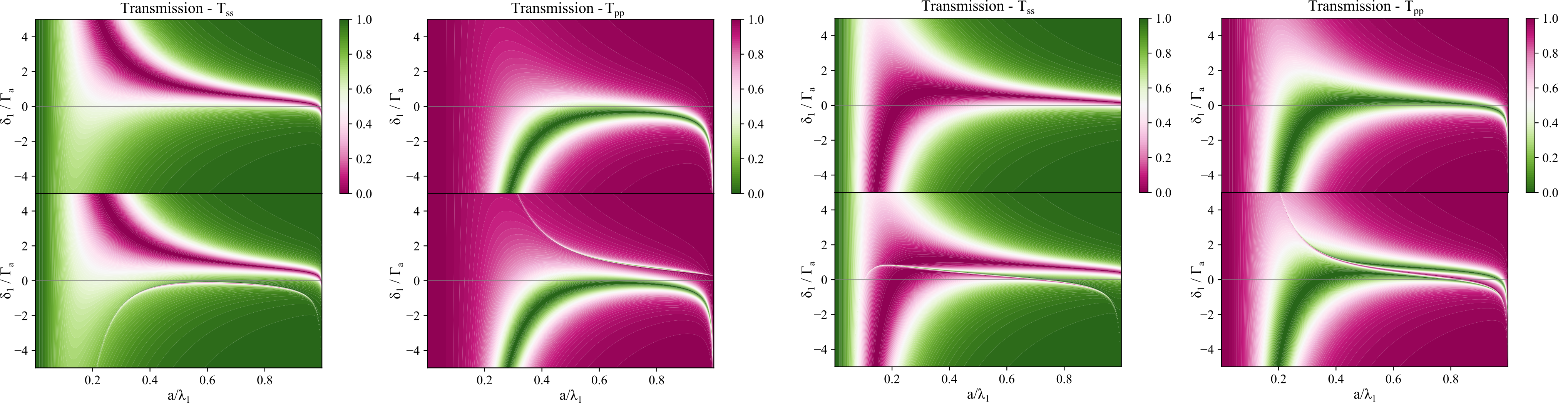}
  \caption{Transmission spectra ($T_{ss}$ and $T_{pp}$) as function of detuning $\delta_1/\Gamma_a$ and lattice period $a/\lambda_{1}$. In the top panels the emitters are identical, $\omega_2 = \omega_1 $, while in the bottom panels $\omega_2 = \omega_1 + \Gamma_a /2$. The two left columns correspond to a lattice with basis vector $\textbf{T}=(0,0.6)a$, and the two right columns to $\textbf{T}=(0.7,0.5)a$. }
  \label{SMtransmission}
\end{figure}

Fig.\ref{SMtransmission} shows transmittance spectra for $s-$ and $p-$polarizations as function of detuning and lattice periocity for the two lattices considered in Fig. 5 of the main text. 

\begin{figure}[ht]
  \centering
  \includegraphics[width=0.9\textwidth]{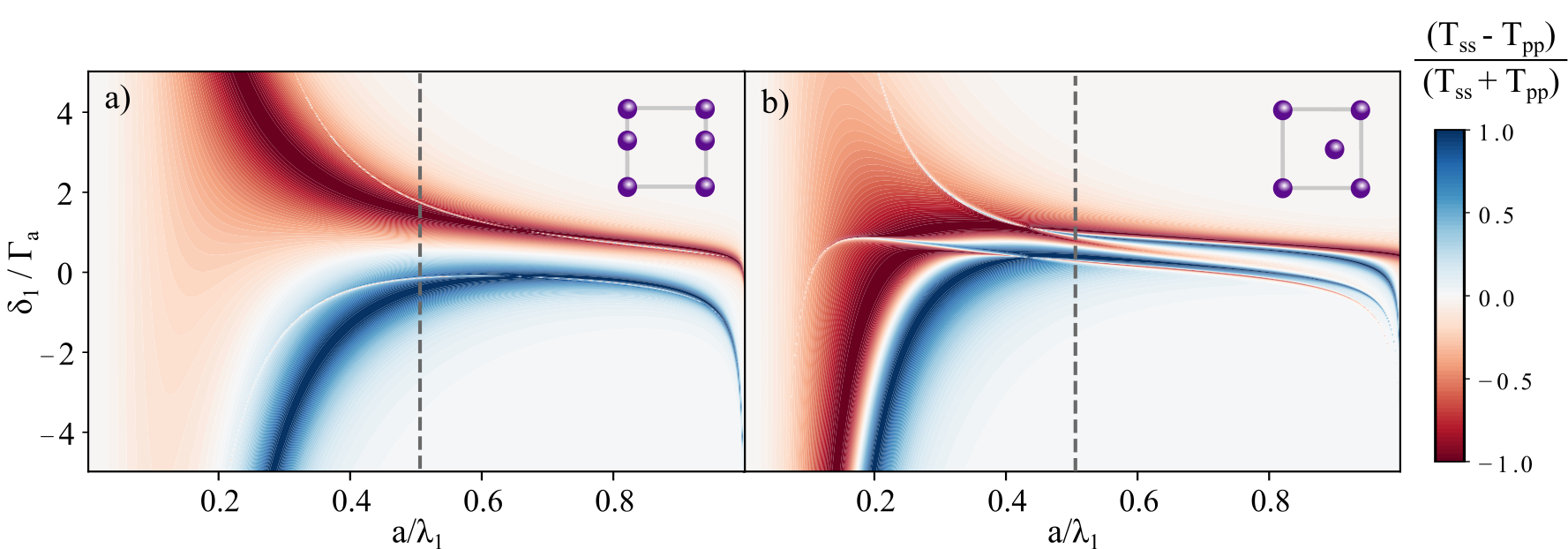}
  \caption{Visibility of transmission components $(T_{ss}-T_{pp})/(T_{ss}+T_{pp})$ as function of detuning $\delta_1/\Gamma_a$ and lattice period $a/\lambda_{1}$ with detuned emitters, $\omega_2 = \omega_1 + \Gamma_a /2$. In (a) $\textbf{T}=(0,0.6)a$ and in (b) $\textbf{T}=(0.7,0.5)a$. }
  \label{SMtrans}
\end{figure}

In Fig.~\ref{SMtrans} we show the visibility of  transmission components $(T_{ss}-T_{pp})/(T_{ss}+T_{pp})$ at normal incidence for the two lattices used in the main text as polarizers, but with detuned emitters, $\omega_2= \omega_1 + \Gamma_a /2$. The contour plots show the efficiency of the polarizer effect as function of the lattice period and detuning. The quasi-BICs that arise with frequency detuning result in the emergence of very narrow features. 
The dashed vertical line represents the period used for the plots shown in 
Fig.~5 of the main text.
\bibliographystyle{apsrev4-1}
\bibliography{references}